\documentclass[conference]{IEEEtran}
\IEEEoverridecommandlockouts
\usepackage{cite}
\usepackage{amsmath,amssymb,amsfonts}
\usepackage{algpseudocode}
\usepackage{graphicx}
\usepackage{textcomp}
\usepackage{xcolor}

\def\BibTeX{{\rm B\kern-.05em{\sc i\kern-.025em b}\kern-.08em
    T\kern-.1667em\lower.7ex\hbox{E}\kern-.125emX}}
\usepackage[normalem]{ulem}
\usepackage[acronym]{glossaries}
\usepackage{soul} 
\usepackage{tabularx}
\usepackage{tcolorbox}
\tcbuselibrary{skins,breakable}

\usepackage[htt]{hyphenat} 
\newcolumntype{P}[1]{>{\raggedright\arraybackslash}p{#1}}
\newcommand{\cell}[1]{\begin{minipage}[t]{\linewidth}#1\end{minipage}}

\usepackage{graphicx}
\usepackage{textcomp}
\usepackage{colortbl}
\usepackage{flushend} 
\usepackage{float}
\usepackage{subcaption}
\usepackage{multirow}
\usepackage{multicol}
\usepackage{xspace}
\usepackage{comment}
\usepackage[normalem]{ulem}
\usepackage{multirow}
\usepackage{stfloats} 
\usepackage{url}
\usepackage{longtable,booktabs}
\usepackage[version=4]{mhchem}
\usepackage[compact]{titlesec}
\usepackage{setspace}
\usepackage[dvipsnames]{xcolor}
\usepackage{annotate-equations}
\usepackage{mathrsfs}
\usepackage{algorithm}
\usepackage{algpseudocode}
\def\secformat{\bfseries}
\usepackage{listings}
\usepackage{pifont}

\newtcolorbox[auto counter]{finding}[1][]{%
    colback=blue!5,           
    colframe=blue!40,         
    boxrule=0pt,              
    leftrule=2mm,             
    sharp corners,            
    before upper={\textbf{Finding~\thetcbcounter:}~}, 
    fontupper=\normalfont,    
}

\definecolor{neongreen}{rgb}{0.0, 1.0, 0.0} 
\definecolor{neonpink}{rgb}{1.0, 0.07, 0.58} 
\sethlcolor{neonpink}

\definecolor{teal}{HTML}{0F766E}

\usepackage{enumitem}

\definecolor{secblue}{HTML}{00008B}
\definecolor{subsecblue}{HTML}{0000FF}
\definecolor{subsubsecblue}{HTML}{0047AB}
\definecolor{captionblue}{HTML}{0000FF}
\definecolor{tableheader}{HTML}{203864}
\definecolor{paragraph}{HTML}{3B5E7F}
\definecolor{pgtext}{HTML}{0000CD}
\definecolor{hdrgray}{HTML}{BFCDDB}%
\definecolor{ltgray}{HTML}{DCDCDC}%
\definecolor{cellcolor}{HTML}{BFCDDB}

\titleformat{\subsection}{\secformat\color{subsecblue}}{\thesubsection}{0.5em}{}
\titleformat{\subsubsection}{\secformat\color{subsubsecblue}}{\thesubsubsection}{0.5em}{}

\titlespacing{\parameterization}{0em}{0em}{0em}
\titlespacing{\section}{0em}{0em}{0em}
\titlespacing{\subsection}{0em}{0em}{0em}
\titlespacing{\subsubsection}{0em}{0em}{0em}
\titlespacing{\paragraph}{0em}{0em}{0em}
\glsdisablehyper
\makeglossaries   
\newacronym{tl}{TL}{Transfer Learning}
\newacronym{hpc}{HPC}{High Performance Computing}
\newacronym{ml}{ML}{Machine Learning}
\newacronym{nlp}{NLP}{Natural Language Processing}
\newacronym{amm}{AMM}{Abstract Machine Model}
\newacronym{sipt}{M\textsubscript{$\mathcal{SA}$}}{\texttt{Stacked Input Alignment Model}}
\newacronym{ipt}{M\textsubscript{$\mathcal{A}$}}{\texttt{Input Alignment Model}}

\newacronym{cn}{$\mathcal{M}odel\mathcal{X}$}{\texttt{Cross Prediction Model}}
\newacronym{mse}{MSE}{Mean Square Error}
\newacronym{nn}{NN}{Neural Network}
\newacronym{lp}{LP}{Linear Probing}
\newacronym{ft}{FT}{Fine-tuning}
\newacronym{ind}{IND}{In-Distribution}
\newacronym{ood}{OOD}{Out-of-Distribution}
\newacronym{fsl}{FSL}{Few-Shot Learning}
\newacronym{sssp}{SSSP}{Single-Source Shortest Path}
\newacronym{ae}{AE}{Auto Encoder}
\newacronym{vae}{VAE}{Variational Auto Encoder}
\newacronym{tab}{TabNet}{Transformer}
\newacronym{llm}{LLM}{Large Language Model}
\newacronym{llms}{LLMs}{Large Language Models}

\definecolor{sg}{HTML}{45B39D}

\definecolor{sgl}{HTML}{A2D9CE}
\definecolor{bl}{HTML}{5499C7}
\definecolor{bll}{HTML}{A9CCE3}
\definecolor{or}{HTML}{D35400}
\definecolor{orl}{HTML}{EDBB99}
\definecolor{usecase}{HTML}{008FAC}
\definecolor{deepskyblue}{HTML}{00BFFF}
\definecolor{purple}{HTML}{C4A9FF}
\definecolor{bb}{HTML}{DAE3F3}
\definecolor{oo}{HTML}{FBE5D6}
\definecolor{gry}{HTML}{EDEDED}
\definecolor{yl}{HTML}{FFC000}
\definecolor{yll}{HTML}{FFFFE0}

\newcommand{\tablecellbl}[1]{\cellcolor{bll}\textcolor{black}{#1}}
\newcommand{\tablecellor}[1]{\cellcolor{orl}\textcolor{black}{#1}}
\newcommand{\tablecellgr}[1]{\cellcolor{sgl}\textcolor{black}{#1}}

\newcommand{\tablecellhl}[1]{\cellcolor{blue!15}\textcolor{black}{#1}}

\usepackage{tikz}

\definecolor{green}{rgb}{0.1,0.1,0.1}

\newcommand{\sysname}{\textsc{Compass}\xspace}

\newcommand{\bofire}{\texttt{BoFire}\xspace}

\newcommand{\dice}{\texttt{DiCE}\xspace}

\usepackage{diagbox}
\usepackage{fontawesome5}

\newcommand{\begreat}{\texttt{GReaT}\xspace}

\newcommand{\tabcf}{\texttt{TABCF}\xspace}

\usepackage{caption}
\usepackage{dcolumn, array, booktabs}
\usepackage{makecell}
\newcolumntype{.}{D{.}{.}{-1}}

\AtBeginDocument{%
  \providecommand\BibTeX{{%
    Bib\TeX}}}

\newcommand{\datasetcount}{8\xspace}

\newcommand{\reco}{\texttt{\textbf{recommend}}\xspace}
\newcommand{\modify}{\texttt{\textbf{reconfigure}}\xspace}
\newcommand{\whatif}{\texttt{\textbf{what-if}}\xspace}

\newcommand{\largesize}{126GB\xspace}
\newcommand{\bgreat}{\textbf{GReaT}}

\makeatletter
\newcommand{\linebreakand}{%
  \end{@IEEEauthorhalign}
  \hfill\mbox{}\par
  \mbox{}\hfill\begin{@IEEEauthorhalign}
}
\makeatother

\begin{document}
\pagenumbering{arabic}
\pagestyle{plain}
\title{\sysname: A Unified Decision-Intelligence System for Navigating Performance Trade-off in HPC}


\author{
  \IEEEauthorblockN{Ankur Lahiry}
  \IEEEauthorblockA{
    \textit{Texas State University} \\
    San Marcos, TX, USA
  }
  \and
  \IEEEauthorblockN{Banooqa Banday}
  \IEEEauthorblockA{
    \textit{Texas State University} \\
    San Marcos, TX, USA
  }
  \and
  \IEEEauthorblockN{Yugesh Bhattarai}
  \IEEEauthorblockA{
    \textit{Texas State University} \\
    San Marcos, TX, USA
  }
  \linebreakand
  \IEEEauthorblockN{Mohammad Zaeed}
  \IEEEauthorblockA{
    \textit{Texas State University} \\
    San Marcos, TX, USA
  }
  \and 
  \IEEEauthorblockN{Tanzima Z. Islam}
  \IEEEauthorblockA{
    \textit{Texas State University} \\
    San Marcos, TX, USA
  }

}

\maketitle
\pagenumbering{arabic}
\pagestyle{plain}
\begin{abstract}

HPC systems expose many configuration parameters that jointly drive competing objectives. Existing tools such as autotuners recommend good configurations but do not identify minimal changes for a near-miss configuration to meet a performance objective, and they often ignore domain-specific constraints. To address this gap, we introduce \sysname---a modular, programmable engine that uses operational traces to generate HPC configuration recommendations and guide tuning decisions. This paper: (1) formalizes configuration questions into query patterns; (2) develops an interactive decision-making engine that formulates these queries as \gls{ml} tasks; 
(3) quantifies the trustworthiness of its recommendations by providing evidence and quantifying uncertainty, and---when confidence is low---provides guidance on which configurations to run next. We validate \sysname using analytical ground truth, reconstruction accuracy, reproduction of published findings, and when possible, running on real hardware. When integrated with an open-source HPC scheduling simulator, \sysname cuts average job turnaround time by 65.93\% and node usage by 80.93\% relative to the state-of-the-art. Moreover, \sysname achieves up to 100× faster training and 
80× faster inference than state-of-the-art generative methods, and scales to traces with 1.3B samples and 126GB of data.

\end{abstract}



\section{Introduction}



\gls{hpc} systems drive discoveries in materials design, genomics, astrophysics, and national security by running simulations and analyses that conventional platforms cannot. These systems combine CPUs, GPUs, and other accelerators, so the same workload can experience different performance, power, and resource usage trade-offs depending on how it is configured to run. Moreover, different stakeholders pursue different objectives: users want low time-to-solution, system administrators must balance fairness, efficiency, and power constraints, and facility operators focus on maximizing throughput while controlling energy costs. 

\textbf{Motivation.} These goals conflict in practice. For instance, lowering power caps can reduce energy use but increase runtime for compute-bound jobs. 
Together, these examples show that configuration selection is a multi-objective decision-making problem that is both critical and hard. These observations motivate our work: the HPC community needs a data-driven decision-making engine that can automatically guide users toward a desirable solution.

\textbf{Limitations of Existing Tools and Methods.} 
Existing \gls{hpc} tools do not provide the kind of decision support users require. For instance, most optimization and autotuning tools search for one fixed objective and return one configuration~\cite{ansel2014opentuner,liu2021gptune,tapus2002active,durholt2025bofire,park2025hyperf} without considering domain knowledge. To our knowledge, no framework exists that jointly support the three recurring user questions we study: what configuration to start with, how to change a current run, and what will happen if a parameter changes. As explanation, these tools report uncertainty, but not the evidence for why a suggestion is made or what to do when no supported suggestion can be made. 

Moreover, existing data-driven methods solve only one part of the configuration selection problem at a time. For instance, existing generative modeling methods can synthesize plausible configurations, but they do not tell a user how to modify a specific prior run. Counterfactual methods can aid a user to reconfigure a prior under performing configuration, yet they do not provide domain experts to specify constraints that cannot be violated. For example, the A1 partition of Marconi 100 provides 4 NVIDIA V100 GPUs per node, so requesting 16 GPUs without enforcing a minimum of 4 nodes yields a valid-looking yet unexecutable configuration. Predictive models can estimate what will happen for a given configuration, but they cannot generate new feasible alternatives. Thus, while each method answers one question, none jointly supports recommendation, reconfiguration, and what-if reasoning. 

\textbf{Contributions.} To address these gaps, we design \sysname---an interactive system for adaptive configuration selection. \sysname guides users, system software, and schedulers to choose feasible configurations under competing objectives, while allowing them to specify constraints that must be satisfied. 
Concretely, we make the following contributions.
\textbf{First,} to replace ad hoc, trial-and-error-based configuration selection, we draw on Human-Computer Interaction (HCI) principles for question design~\cite{amershi2019guidelines,doshi2017towards,kules2008designing} and a survey of over 100 HPC papers (Section ~\ref{sec:query-template})
to organize the three recurring user questions identified above into three templates: \reco, \modify, and \whatif. 
Specifically, \reco answers what configuration to start with, \modify answers how to change a current run, and \whatif answers what will happen if a parameter changes. Mathematically, these queries map to 3 distinct ML tasks: \reco$\rightarrow$inverse modeling, \modify $\rightarrow$ counterfactual generation, and \whatif $\rightarrow$ forward prediction. 

\textbf{Second,} to support all three queries within one system, we formulate them as a Conditional Constrained Counterfactual Generation ($C^3G$) task. Treating these pieces separately would require different models for prediction, generation, and constraint checking, in addition to extra rules to prevent them from producing conflicting results. In contrast, the \(C^3G\) loss (Section~\ref{sec:modeling-engine}) encodes four requirements in one objective: (i) \textbf{validity}: so the output meets the target; (ii) \textbf{implementability}: so it stays close to the user's baseline; (iii) \textbf{feasibility}: so it respects user-specified constraints; and (iv) \textbf{diversity}: so it returns multiple distinct good-enough alternatives. 
\textbf{Third,} in addition to generating actionable configurations, we propose Algorithm~\ref{alg:assess_reliability} to assess how trustworthy each suggestion is 
and provide the evidence and guidance that existing tools lack. Specifically, it attaches (1) a label such as \textit{trusted}, \textit{caution}, or \textit{unsupported}; (2) the number and identities of observed samples that support the suggestion; and (3) when no supporting data exists, an explicit indication that this region of the space is unsupported plus a small set of configurations to run next 
to improve the model for similar future queries.

\textbf{Finally,} we implement \sysname as a modular, distributed-memory system for large performance logs. The pipeline (1) collects query requirements via an AI chat bot, (2) uses a Dask-based dataframe layer for parallel file reads and information-aware sampling~\cite{pmlr-v28-mineiro13} to reduce multi-terabyte traces to an informative subset, (3) either trains a surrogate model on this subset or uses a user-provided one, (4) minimizes the $C^3G$ loss to synthesize the top-$\gamma$ configurations (user-chosen $\gamma$), and (5) returns an actionable outcome with a trustworthiness label and its supporting samples. \sysname exposes explicit interfaces for surrogate models, data reduction, domain constraints, and pre-processing, so users can swap in their own components and trace loaders without modifying the rest of the pipeline. 
The current prototype of \sysname supports performance profiles stored as single- or multi-run files as well as HPCToolkit~\cite{HPCToolkit} traces.
\textbf{To our knowledge, \sysname is the first interactive decision-support system that recommends an initial configuration when none is known and then supports iterative refinement through reconfiguration queries.}

\textbf{Validation.} Validation across three axes on \datasetcount datasets from production HPC systems show that: (1) when run on real hardware in a multi-node benchmark, \sysname's trusted configurations achieve execution times within about 7\% of \sysname's predictions. (2) on held-out test samples, when query-relevant fields are masked, \sysname recovers the original hidden values to within about 1\% error on average; and (3) \sysname matches analytical ground truth found in the literature and reproduces prior configurations and findings to within about 1\% error on average, often using only 5\% of the data. For example, it reproduces the result in~\cite{patki2019performance} that switching FT’s task binding from ``random'' to ``packed'' at fixed node count reduces run-to-run variability. In a validated scheduler simulator on production traces~\cite{ai-scheduler,dey2025modelx}, its recommendations improve average job turnaround time and system throughput by 65.93\% and 80.93\%, respectively, over a state-of-the-art ML-based scheduler. Overall, COMPASS
reduces inference time by roughly two orders of magnitude
while also avoiding the multi-hour training cost required
by SOTA frameworks.

In summary, we make the following contributions:
\begin{itemize}[noitemsep, topsep=0pt, leftmargin=*]
    \item \textbf{Query Taxonomy:} Formalize HPC decision-intelligence queries into 3 design patterns---\reco, \modify, and \whatif.
    \item \textbf{Unified Generative Loss:} Propose a new loss function $C^3G$ to unify these three queries.
\item \textbf{Interpretable Trustworthiness Assessment with Mitigation:} Introduce Algorithm~\ref{alg:assess_reliability}, which explains why a configuration is suggested, quantifies its trustworthiness, and identifies what additional data would improve confidence.
\item \textbf{Rigorous Validation:} Validates across four complementary axes---reconstruction accuracy, analytical ground truth matching, reproduce knowledge, and run on real hardware.
\item \textbf{Demonstrate Benefit for Job Scheduler:} Demonstrate using a validated scheduler simulator that \sysname's recommendations improve average job turnaround time and system-wide throughput compared to published results.
\item \textbf{Scale to \largesize Trace:} Implement \sysname as a public, modular framework that scales to massive traces with up to \(100\times\) faster training, \(80\times\) faster inference, and higher reconstruction accuracy than state-of-the-art generative models.
\end{itemize}

\textbf{Limitations \& Scope.} \sysname is not a replacement for domain expertise or real-time schedulers.
It can act as a one-shot advisor or run in an interactive loop where users can ingest data from new runs so its suggestions can adapt to performance drift and workload evolution.
Although our survey also identified a query type targeting the discovery of performance bottlenecks, this query does not yield actionable configuration suggestion, therefore, it does not align with the objective of this work and we defer it for future.

\section{Taxonomy of HPC Queries}
\label{sec:query-template}
\textbf{To replace today's ad hoc, trial-and-error-based configuration selection process with a principled data-driven approach, guided by principles from human–computer interaction and explainable AI~\cite{amershi2019guidelines, doshi2017towards, kules2008designing}, we organize \gls{hpc} decision-support queries into three distinct and conceptually non-overlapping query types.} 
The three query types are not just different ways to ask the same user question; they require different kinds of ML tasks. Two of them involve generation (constrained and counterfactual) under constraints, while one is prediction. 
Table~\ref{tab:query-templates} summarizes these three query types. To validate this taxonomy, we systematically surveyed over $100$ HPC decision-support papers published over the past 30 years (1995 - 2025) ~\cite{mu_utilization_2001, schroeder_large-scale_2010, smith_using_1999, feitelson_job_1995, Backfilling1, feitelson_metrics_1998, SchedMDSlurm, lublin_workload_2003, lee_are_2005, feitelson_experimental_2005, gaussier_improving_2015, feitelson1, feitelson2, Backfilling5, ParallelJobSched, liang_failure_2007, yuan_job_2012, feitelson_experience_2014, carastan-santos_obtaining_2017, fan_trade-off_2017, zhang_rlscheduler:_2020, tanash2019high-performance, patki2019performance, simakov_slurm_2018, tanash_ensemble_2021, fan_deep_2021, zhang_schedinspector:_2022, brown_predicting_2022, gadban_analyzing_2022, gadban_investigating_2020, gadban_analyzing_2021, kolker-hicks_reinforcement_2024, menear_mastering_2023, antici2023pm100, antici2024f, chu2023hotcloudperf, chu2023how, hariharan_end-to-end_2020, reuther_interactive_2024, león-vega_comprehensive_2024, cornelius2025extracting, jakobsche_autonomy_2025, abdurahman_scalable_2025, menear_tandem_2025, gaikwad_predictive_2025, mckerracher2025fresco, Loi2025Uncertainty, menear2024high, reuther_scalable_2018, cunha_job_2017,silva_jobpruner:_2018,kunkel_understanding_2018, samsi2021mitsuperclouddataset, witt_feedback-based_2019, kintsakis_reinforcement_2019, burns_borg_2016, lehmann_how_2023, farnes_building_2018, tovar_job_2017, chiang_impact_2002, smith_predicting_2004, andresen_machine_2018, berral_towards_2010, Yang:sc13, ZhouBGQ, PatkiTechRep, wallace_data_2016,saillant_predicting_2020, palmer_open_2015,agarwal_page_2015, allen_demystifying_2021, allen_in-depth_2021, bachkaniwala_lotus:_2024, bertolli_performance_2024, che_rodinia:_2009, chen_porple:_2014, chien_performance_2019, choi_memory_2022, cooper_shared_2024, fusco_understanding_2024, ganguly_interplay_2019, landaverde_investigation_2014, lin_drgpum:_2023, nataraja_enhanced_2024, schieffer_harnessing_2024, schieffer_inter-apu_2025, shen_cudaadvisor:_2018, tandon_porting_2024, dazzi_intelligent_2024, lopez_lessons_2021, lehong_towards_2015, simakov_developing_2022, scully-allison_same_2024, menear_tandem_2024, antici_online_2023, dazzi_intelligent_2023, jiang_hpc_2021, rambler_chaid_2010, witt_predictive_2019, wang_novel_2019, lockwood_2018_1345780}, identifying the core research questions in each work and categorizing them under the proposed taxonomy. 
This review confirms the long-term evolution of the field from early scheduling heuristics to modern AI- and \gls{ml}-driven policies. From each paper, we extract individual research questions and map each to exactly one query category---\reco, \modify, and \whatif. While a single paper may span multiple tasks, we map each research question to a unique query pattern. 
\begin{table}[t]
\tiny
\caption{Summary of query templates prevalent in HPC decision-support scenarios.}
\label{tab:query-templates}
\setlength{\tabcolsep}{4pt}
\renewcommand{\arraystretch}{0.95}
\resizebox{\columnwidth}{!}{%
\begin{tabular}{|p{1.2cm}|p{4.95cm}|}
\hline

\rowcolor{bll}
\multicolumn{2}{|c|}{\rule{0pt}{1.9ex}\textbf{\reco. ML Task: Inverse Problem}\rule[-0.6ex]{0pt}{0pt}}\\ \hline

\textbf{Intent \& HPC Scenarios:} &
Given target objectives $Y$, find values of $\mathcal{X}_{\text{user}}$ to meet those objectives on given optional constraints $C$. Use cases: Novice HPC users; Software systems such as job schedulers, resource managers, HPC facilities.
\\ \hline

\textbf{Template:} &
\textcolor{blue}{Recommend a configuration} $\mathcal{X}_{\text{user}} = \{x_i : v_i, x_i : ?\}$
\textcolor{blue}{to achieve} $Y=\{y_1: \mathcal{O}_{1}, \dots, y_m : \mathcal{O}_{m}\}$
\textcolor{blue}{where} $C$ = $\{\!x_j\!:\!v_j\!\}$
\\ \hline

\textbf{User provides:} &
Objective(s) over $Y$ and constraints $C$, specifying configurable parameters in $\mathcal{X}_{user}$, known values for these can be specified or left unspecified ' while leaving remaining parameters unspecified.
\sysname responds: Candidate configurations for $\mathcal{X}_{user}$, generated by conditioning on the empirical distribution of $\mathcal{X}_{sys}$ observed from historical jobs that satisfy $C$.
\\ \hline

\rowcolor{sgl}
\multicolumn{2}{|c|}{\rule{0pt}{1.9ex}\textbf{\modify. ML Task: Counterfactual Reasoning.}\rule[-0.6ex]{0pt}{0pt}}\\ \hline

\textbf{Intent \& HPC Scenarios:} &
Given desired outcomes $Y$ and an initial configuration, update existing $\mathcal{X}_{user}$ with minimal changes. Use cases: Autotuners; Performance tuning.
\\ \hline

\textbf{Template:} &
\textcolor{blue}{Reconfigure}
$\mathcal{X}_{\text{user}} =
\{x_1:v_1^{old}\rightarrow~v_1^{new},\!\dots\!, x_n:\!v_n^{old}\!\rightarrow\!?\}$
\textcolor{blue}{to achieve} $Y$ = $\{y_j:v_j^{old}\rightarrow\!v_j^{new}\}$
\textcolor{blue}{where} $C$ = $\{x_j\!:\!v_j\}$
\\ \hline

\textbf{User provides:} &
A baseline configuration
$x \in \{\mathcal{X}_{user} \cup \mathcal{X}_{sys}\}$,
desired changes in $Y$, and optional constraints $C$.
\textbf{\sysname responds:} Configurations that minimally modify
$\mathcal{X}_{user}$ while satisfying $C$; all unspecified features
are initialized to their baseline values but may change if required by
feature correlations, in which case \sysname returns the closest feasible
alternatives and defers the final decision to the user.
\\ \hline

\rowcolor{orl}
\multicolumn{2}{|c|}{\rule{0pt}{1.9ex}\textbf{\whatif. ML Task: Forward Problem.}\rule[-0.6ex]{0pt}{0pt}}\\ \hline

\textbf{Intent \& HPC Scenarios:} &
Given a change in $\mathcal{X}_{user}$ while keeping $\mathcal{X}_{sys}$ unchanged, estimate its effect on outcomes $Y$ (Forward problem).
Users or software systems decides based on the impact of changing one or more inputs on a desired outcome.
\\ \hline

\textbf{Template:} &
\textcolor{blue}{What-if we change}
$\mathcal{X}_{\text{user}} = \{x_1: v_1^{old} \rightarrow v_1^{new}, \dots, x_n: v_n^{old} \rightarrow v_n^{new}\}$,
\textcolor{blue}{how would that affect}
$Y = \{y_1, \dots, y_m\}$
\textcolor{blue}{where} $C$ = $\{x_j\!:\!v_j\}$
\\ \hline

\textbf{User provides:} &
A baseline configuration $X \in \{\mathcal{X}_{user} \cup \mathcal{X}_{sys}\}$, desired changes to $\mathcal{X}_{user}$, and constraints $C$, with all unspecified features set to their baseline values to form a complete configuration.
\sysname responds: Predicted changes in outcomes $Y$.
\\ \hline

\end{tabular}%
}
\end{table}
\subsection{Data Description}
\label{sec:data-description}
\textbf{Features.}
As discussed in Section~\ref{sec:query-template}, each query is formulated over a dataset
$D = \{(\mathcal{X}_i, Y_i)\}_{i=1}^N$, where $\mathcal{X}_i \in \mathbb{R}^d$ denotes a feature
vector. Here, $N$ is the total number of samples in $D$, there are $d$ features supporting both numerical and categorical, and $m$ target objectives. $D$ represents dataset with each row corresponding to a sample and each column corresponding to a feature or a target. 
Without loss of generality, we partition $\mathcal{X}$ into user-controllable parameters $\mathcal{X}_{\text{user}}$ and externally determined parameters $\mathcal{X}_{\text{sys}}$, which are fixed by the system environment (e.g., hardware counters, scheduler, or facility policies) and not directly adjustable by the user.
Users can encode domain knowledge in the constraint list $C$.

\noindent
\textbf{Targets.} \sysname supports multiple targets of both regression and classification types denoted by $Y_i \in \mathcal{Y}$. Here,
$\mathcal{Y} \subseteq \mathbb{R}^m$ for regression, capturing performance outcomes
such as \texttt{runtime}, \texttt{energy} or \texttt{throughput}, while for classification $\mathcal{Y}$ is a
finite set of categorical labels, such as \texttt{is\_anomaly} or \texttt{job\_state}. Targets are not controllable; they are observed outcomes resulting from a given configuration. 

\subsection{Recommendation Query}
\label{sec:reco}
A \reco query addresses goal-oriented decision-making by asking how to configure a job to achieve specified performance objectives. Formally, it is an \textit{inverse problem}: users define desired outcomes $Y$, specify known configuration values $v_i$ in $\mathcal{X}_{\text{user}}$, leave remaining parameters unspecified, and provide constraints $C$ that encode domain knowledge—such as bounds, equalities, inequalities, or functional relationships among parameters. This partial specification separates knowns from unknowns, encodes domain knowledge through $\mathcal{X}{\text{user}}$, and determines unspecified parameters by conditioning on observed data subject to $C$. In HPC settings, \reco queries commonly arise for novice users configuring job submissions, for self-configuration in system software, and for resource provisioning under performance goals. 

\subsection{Reconfiguration Query}
\label{sec:reconfig}
The \textit{intent} of this query is to refine a known configuration that does not meet a desired outcome. Unlike a \reco query, which operates without an initial state, a \modify query begins from an existing setup. It asks what specific changes to controllable parameters will achieve a target shift in outputs while preserving system constraints. This represents a \textit{constrained counterfactual reasoning task}. Table~\ref{tab:query-templates} provides the query template.
Unlike \reco, all parameters not included in $X_{user}$ are preserved from the baseline by default, but may vary when required to satisfy constraints or account for feature dependencies. In this case, \sysname prioritizes constraint satisfaction and minimal deviation and defers final decision to the user. 
In \textit{HPC scenarios}, this query is used by experienced users, auto-tuning systems, and adaptive system software to refine configurations near known operating points.

\subsection{What-if Exploration Query}
\label{sec:what-if}
The intent of this query is to understand: ``Given the configuration, if I just change this parameter, but want to keep Y the same, what else would have to change?" 
Table~\ref{sec:query-template} provides the template.
This query represents a \textit{forward problem}.
Parameters not explicitly included in $X_{\text{user}}$ or constrained by $C$ are initialized to their baseline values to form a complete configuration, and may vary only when required to satisfy constraints or account for feature dependencies. In \textit{HPC scenarios}, this query is common in scheduling systems, energy-aware runtime management, or dynamic tuning, where speculative reasoning helps avoid costly trial runs. \textit{Users provide} the current configuration and change in values for one or more input features. \sysname's \textit{output} is the predicted effect on the target variables, based on $f_{\theta}$.

\section{Our Approach}
\label{sec:methodology}
\begin{figure}
    \centering
    \includegraphics[width=\linewidth]{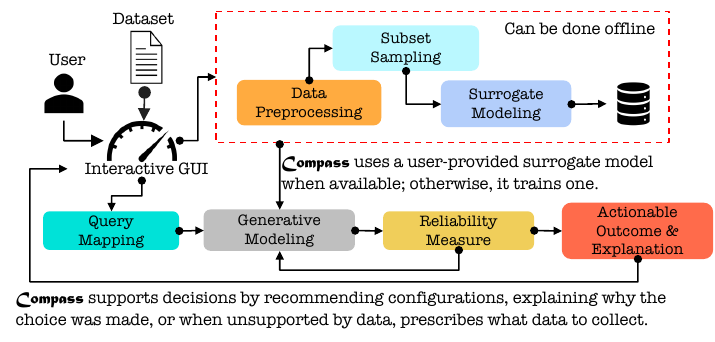}
    \caption{Overview of \sysname. \sysname supports users by taking a dataset and query, mapping the request into a formal decision task, generating feasible answers, assessing their reliability, and returning an actionable recommendation with explanation. If the user provides a pre-trained performance model, it is used directly as the surrogate; otherwise, \sysname constructs one using subset sampling and surrogate modeling, optionally offline}
    \label{fig:overview}
\end{figure}

Figure~\ref{fig:overview} shows the end-to-end workflow of \sysname. \sysname comprises of a front-end and back-end. Through the front-end chat bot, users upload data, enter queries, and optionally upload a pre-trained model. 
Implementation details of the rest of the components are 
discussed in Section~\ref{sec:implementation}.

\subsection{Surrogate Modeling}
\label{sec:surrogate-modeling}
\label{subsection:surrogate_modeling}
While configuration generation is a generative task, pure generative modeling is data intensive: it requires extensive labeled data and increases training time. 
Instead, \sysname leverages a surrogate model that represents the decision boundary and evaluates how far each candidate is from that boundary. This principle of self-consistency makes the configuration generation process time- and data-efficient. 
In \sysname, surrogate performance modeling is a plug-and-play component. Rather than developing a new predictor, \sysname builds on the large body of existing work on predictive modeling for system performance~\cite{islam2016a, thiagarajan2018bootstrapping, thiagarajan2018paddle, jain2013predicting, witt_predictive_2019, patki2019performance, nabi2022optimaltrainingfairpredictive, guerra2023costoftraining, 10460066, phelps2024reimagine}. 
This design explicitly decouples decision reasoning from the specific model ($f_{\theta}$) used. This means \sysname can work with any predictive model, no matter how it was built.

\textbf{Scenario A: User provides a pre-trained model.} If a user provides a pre-trained model, \sysname uses it directly. The model architecture and weights can be uploaded as a serialized file, such as \texttt{.h5} or \texttt{.pt}, through the front-end chatbot.
\textbf{Scenario B: No model is provided.} If no pre-trained model is available, \sysname trains one from the uploaded data by splitting 
$D$ into training (80\%) and validation (20\%) sets, training multiple candidate models in parallel, and selecting the model with the lowest validation error. The selected model is then saved and can be reused in later sessions. Table~\ref{tab:all_models_results} summarizes the surrogate modeling approaches currently supported by \sysname.
\textbf{Why \sysname is Robust.} 
When new data become available, \sysname can update the surrogate model. In the current system, this is done by retraining from scratch. More efficient adaptation mechanisms, such as transfer learning or fine-tuning, are left for future work because they are engineering extensions rather than methodological contributions of this paper.
While complex surrogate model architecture can improve statistical confidence of the generated configurations, they do not alter the core backend engine. Thus, building a robust surrogate model is \textit{orthogonal to \sysname's contributions}. 
%
\subsection{Unified $C^3G$ Loss}
\label{sec:modeling-engine}
\label{sec:cf-backbone}
\textbf{The second methodological contribution of this paper is a generative modeling formulation that unifies the three query types in Section~\ref{sec:query-template} under a single loss function.}
Unlike predictive models, that estimate the outcome of a given configuration, our task starts with a desired outcome and must construct one or more configurations that achieve it. 
This is challenging because performance target values are often close to one another, so nearby objective values can map to multiple similar configurations. As a result, the objective function alone may not distinguish which configuration should be selected.

To address this challenge, we formulate configuration generation as a \emph{conditional constrained counterfactual generation} task, denoted \(C^3G\), where \emph{conditional} refers to the user-specified objective, \emph{constrained} enforces hard domain rules, and \emph{counterfactual} means generation starts from a baseline configuration and modifies it toward the target. Equation~\ref{eqn:c3g_cf_loss} then defines a loss function that encodes four elements: (i) the surrogate model \(f_{\theta}\), which evaluates candidates against the objective; (ii) a baseline configuration \(\mathbf{x}\) drawn from the data; (iii) a hard constraint set \(C\); and (iv) a diversity term that encourages multiple good-enough configurations. Users provide \(\mathbf{x}\) and \(C\) through the front-end chatbot.
\begin{figure*}[t]
\tiny
\setlength{\tabcolsep}{25pt} 
\begin{tabular}{p{0.9\textwidth}}
\begin{minipage}{\textwidth}
\begin{equation}
\label{eqn:c3g_cf_loss}
\eqnmarkbox[Green]{cf_loss}{
\{\mathbf{x}'_1,\ldots,\mathbf{x}'_N\}
}
=
\arg\min_{\{\mathbf{x}'_i \in \mathcal{X}_{\mathrm{valid}}\}}
\left[
\lambda_{\mathrm{valid}}\,
\eqnmarkbox[WildStrawberry]{validity}{
\mathcal{L}_{\mathrm{valid}}\!\big(\{\mathbf{x}'_i\},y^{\star},f_{\theta}\big)
}
+
\lambda_{\mathrm{prox}}\,
\eqnmarkbox[YellowOrange]{proximity}{
\sum_{i=1}^d \mathcal{L}_{\mathrm{prox}}(\mathbf{x}'_i,\mathbf{x})
}
+
\lambda_{\mathrm{cons}}\,
\eqnmarkbox[RoyalBlue]{cons}{
\sum_{i=1}^N \Phi(\mathbf{x}'_i)
}
-
\lambda_{\mathrm{div}}\,
\eqnmarkbox[Blue]{diversity}{
\mathcal{L}_{\mathrm{div}}\!\big(\{\mathbf{x}'_i\}\big)
}
\right].
\end{equation}

\annotate[yshift=0.9em]{above,right}{cf_loss}{Configuration set}
\annotate[yshift=-0.9em]{below,left}{validity}{Target satisfaction}
\annotate[yshift=0.8em]{above,left}{proximity}{Minimal change from \(\mathbf{x}\)}
\annotate[yshift=0.8em]{above,right}{cons}{Domain constraints}
\annotate[yshift=-0.9em]{below,left}{diversity}{Distinct recourses}
\end{minipage}
\end{tabular}
\footnotesize
\resizebox{0.8\textwidth}{!}{
\begin{tabular}{p{0.3\textwidth} p{0.2\textwidth} p{0.2\textwidth} p{0.1\textwidth}}
\begin{minipage}{0.1\columnwidth}
\begin{equation}
\label{eqn:valid-loss}
\mathcal{L}_{\mathrm{valid}}\!\big(\{\mathbf{x}'_i\}, y^{\star}, f_{\theta}\big)
=
\frac{1}{N}\sum_{i=1}^{N}
\begin{cases}
\big[y_{\min} - f_{\theta}(\mathbf{x}'_i)\big]_+
+
\big[f_{\theta}(\mathbf{x}'_i) - y_{\max}\big]_+,
& \text{if } f_{\theta} \text{ is a regressor}, \\[0.8em]
\displaystyle
\left[
\max_{c \neq y^{\star}} f_{\theta}(\mathbf{x}'_i)_c
-
f_{\theta}(\mathbf{x}'_i)_{y^{\star}}
\right]_+,
& \text{if } f_{\theta} \text{ is a classifier}.
\end{cases}
\end{equation}
\textbf{Validity}\\
\end{minipage}
&
\begin{minipage}{0.1\columnwidth}
\begin{equation}
\label{eqn:prox-loss}
\mathcal{L}_{\mathrm{prox}}(\mathbf{x}'_i,\mathbf{x})
=
\sum_{j=1}^d w_j\,|x'_{i,j}-x_j|
\end{equation}
\textbf{Proximity}\\
\end{minipage}
&
\begin{minipage}{0.1\columnwidth}
\begin{equation}
\label{eqn:cons-loss}
\Phi(\mathbf{x}'_i)
=
\sum_{k} \phi_k(\mathbf{x}'_i)
\end{equation}
\textbf{Constraints}\\
\end{minipage}
&
\begin{minipage}{0.1\columnwidth}
\begin{equation}
\label{eqn:diversity-loss}
\mathcal{L}_{\mathrm{div}}\!\big(\{\mathbf{x}'_i\}\big)
=
\frac{2}{d(d-1)}
\sum_{1\le i<j\le d}
d_{\mathrm{CF}}(\mathbf{x}'_i,\mathbf{x}'_j)
\end{equation}
\textbf{Diversity}\\
\end{minipage}
\\
\end{tabular}
}
\caption{Unified \(C^3G\) objective (top) and explicit definitions of each loss component (bottom).}
\label{fig:c3g_objective}
\end{figure*}

The first component, the \textbf{validity term} \(\mathcal{L}_{\mathrm{valid}}\) (Equation~\ref{eqn:valid-loss}), measures how well the generated configurations satisfy the target outcome under the surrogate model \(f_{\theta}\). Here, \([z]_+ = \max\{z,0\}\) denotes the hinge operator, which keeps the loss at zero when the target objective is met and penalizes only the amount of violation otherwise. For regression tasks where the target objectives are continuous values, the two hinge terms measure whether the predicted outcome falls below \(y_{\min}\) or above \(y_{\max}\), so the loss is zero inside the acceptable range and increases with distance outside it. For classification, the hinge compares the score of the target class \(y^{\star}\) against the highest competing class score, so the loss is zero only when the target class is ranked first and increases when another class is preferred.
The rationale for the proposed range-hinge-based formulation is that most of the HPC performance metrics are continuous-valued objectives are noisy and can vary run-to-run~\cite{patki2019performance}, so insisting on an exact point target (e.g.,
exactly \(5\) seconds) overfits to negligible prediction differences and
artificially narrows the search space. 

The next requirement is \textbf{implementability}: among configurations that satisfy the target, \sysname should prefer ones that require smaller changes from the baseline configuration \(\mathbf{x}\). To encode this requirement, the \textbf{proximity term} \(\mathcal{L}_{\mathrm{prox}}\) (Equation~\ref{eqn:prox-loss}) penalizes large deviations from \(\mathbf{x}\) using a weighted feature-wise distance. The assumption is that this distance is a practical proxy for implementability, so configurations that differ less from the baseline are treated as easier to apply (e.g., adding 4 nodes rather than 1024). The next requirement is \textbf{feasibility}: a configuration is not useful if it violates user-specified domain constraints which could be stating operational rules. Accordingly, the \textbf{domain constraint term} \(\Phi(\mathbf{x}'_i)\) (Equation~\ref{eqn:cons-loss}) aggregates penalties for violating constraints such as hardware compatibility, resource limits, and system-wide policies. This discourages infeasible configurations during optimization rather than removing them after generation. The final requirement is \textbf{choice}: if several configurations satisfy the same target, the system should not return near-duplicates. Therefore, the \textbf{diversity term} \(\mathcal{L}_{\mathrm{div}}\) (Equation~\ref{eqn:diversity-loss}) encourages pairwise separation among generated configurations, so the returned set contains distinct alternatives rather than minor variants of the same solution.

\textbf{Comparison with existing generative losses.}
Existing generative loss functions are designed to reproduce the overall data
distribution (unconstrained), whereas our goal is to generate configurations that satisfy a user-specific target and constraints. For example, Generative Adversarial Networks~\cite{goodfellow2014gan}, CTGANs~\cite{xu2019modeling}, Variational AutoEncoderss~\cite{kingma2014auto}, diffusion models~\cite{ho2020denoising}, and transformer-based models such as Large Language Models~\cite{vaswani2017attention} optimize objectives of the form:
$\min_{\theta}\ \mathcal{L}_{\text{gen}}(\theta)
=
\mathbb{E}_{x\sim p_{\text{data}}}
\big[\ell_{\text{match}}(x;G_{\theta})\big]$,
where \(\ell_{\text{match}}\) is a model-specific distribution-matching loss. Such objectives encourage samples that resemble the training data, but they do not directly encode target satisfaction, user constraints, or proximity to a baseline configuration. Consequently, these requirements are usually enforced after generation through filtering, rejection sampling, or external guidance. In contrast, Equation~\ref{eqn:c3g_cf_loss} encodes them directly in the objective, so generation is driven toward usable configurations. 


\textbf{Comparison with existing counterfactual generation losses.}
Similar to existing counterfactual generation methods, $C^3G$ retains \emph{validity} and \emph{proximity}~\cite{bakir2025dice}. In addition to those two terms, \(C^3G\) also encodes \emph{constraints} and \emph{diversity} and extends the formulation to continuous HPC objectives. As a result, \(C^3G\) supports constrained configuration set generation rather than single-instance outcome change. We extend the \dice library~\cite{bakir2025dice} to implement Equation~\ref{eqn:c3g_cf_loss}.
\subsection{Interpretable Algorithm for Trustworthiness Assessment for Explainability and Mitigation}
\label{sec:reliability}
\label{sec:trustworthiness}
\begin{algorithm}[t]
\footnotesize
\caption{Assessing Reliability of Generated Configurations}
\label{alg:assess_reliability}

\begin{algorithmic}[1]
\Require Candidate configuration $x$, training set $X_{\text{train}}$, validation set $X_{\text{val}}$
\Ensure Reliability label $\in \{\textit{trusted}, \textit{caution}, \textit{unsupported}\}$,
support evidence $\mathsf{support}(x)$, explanation $\mathsf{reason}(x)$

\State \textbf{Compute data support and supporting samples}
\State Fit $k$-NN on $X_{\text{train}}$ 
\State $d(v_i) \leftarrow \frac{1}{k}\sum_{j \in \mathsf{NN}_k(v_i, X_{\text{train}})} \|v_i - x_j\|,\;\forall v_i \in X_{\text{val}}$
\State $\mathsf{NN}(x), d(x) \leftarrow \mathsf{NN}_k(x, X_{\text{train}})$
\State $OOD(x) \leftarrow \frac{1}{N}\sum_{i=1}^{N}\mathbb{I}[d(v_i)\le d(x)]$
\State $\tau_{\text{close}} \leftarrow \text{given}$
\State $\mathsf{support}(x) \leftarrow \{(j,d_j)\in\mathsf{NN}(x): d_j \le \tau_{\text{close}}\}$
\If{$OOD(x) > 0.99$}
    \State $\mathsf{reason}(x)\leftarrow$ ``\textbf{unsupported:} no nearby training samples. Only 1\% of past examples looked this unfamiliar or worse.''
    \State \Return \textit{unsupported}, $\mathsf{support}(x)$, $\mathsf{reason}(x)$
\EndIf
\State \textbf{Compute model uncertainty (UQ)}
\State $u(v_i) \leftarrow \mathrm{Var}\!\left(\{f^{(p)}(v_i)\}_{p=1}^{P}\right),\;\forall v_i \in X_{\text{val}}$
\State $u(x) \leftarrow \mathrm{Var}\!\left(\{f^{(p)}(x)\}_{p=1}^{P}\right)$
\State $UQ(x) \leftarrow \frac{1}{N}\sum_{i=1}^{N}\mathbb{I}[u(v_i)\le u(x)]$
\If{$OOD(x) > 0.95$ \textbf{or} $UQ(x) > 0.95$}
    \State $\mathsf{reason}(x)\leftarrow
    \begin{cases}
    \parbox[t]{0.4\columnwidth}{\textbf{caution:} Limited support samples present.}, &  OOD(x)>0.95\\
    \parbox[t]{0.4\columnwidth}{\textbf{caution:} high model uncertainty despite several support samples present.} & \text{otherwise}
    \end{cases}$
\Else
    \State $\mathsf{reason}(x)\leftarrow$ ``\textbf{trusted:} supported by nearby training samples with stable predictions.''
\EndIf
\State \Return \textit{trusted}, $\mathsf{support}(x)$, $\mathsf{reason}(x)$
\end{algorithmic}
\end{algorithm}

Generated configurations are useful for decision-making only if the system can also indicate how much each suggestion should be trusted. This step is necessary because a generative model can produce configurations that satisfy the target under the surrogate model but lie far from the observed data. Existing HPC performance modeling and autotuning tools typically rely on uncertainty quantification (UQ) alone, using measures such as variance, confidence intervals, or ensemble spread on held-out data. However, UQ reflects model uncertainty; it does not indicate whether a generated configuration is supported by the training distribution. As a result, an out-of-distribution configuration can still receive a confident prediction due to extrapolation, while a configuration in a well-supported region can appear uncertain because of local model sensitivity~\cite{lee2018training, hein2019relunetworksyieldhighconfidence}. \textbf{To address this gap, we introduce a novel interpretable algorithm that acts as a guardrail by scoring each generated suggestion by jointly quantifying predictive uncertainty and data support, and then assigning a label}. 

\subsubsection{Data support via Out-Of-Distribution Detection (OOD)}
\label{sec:ood}
Before interpreting predictive uncertainty, \sysname first checks whether a generated configuration \(x\) is supported by the observed data. Our proposed Algorithm~\ref{alg:assess_reliability} calculates the OOD score using a k-nearest-neighbor (kNN) distance test, independent of the predictive model, following standard OOD detection practice~\cite{gu2019statisticalanalysisnearestneighbor,park2023nearestneighborguidanceoutofdistribution,sun2022outofdistributiondetectiondeepnearest}. First, Algorithm~\ref{alg:assess_reliability} fits a kNN model on the training set \(X_{\text{train}}\) to cluster the training samples and then compute the mean distance to the \(k\) nearest training samples for each validation sample \(v_i \in X_{\text{val}}\) and for the generated configuration \(x\). The validation distances define the typical geometry of observed data, while \(d(x)\) places the generated configuration on the same scale. The out-of-distribution score \(OOD(x)\) is then calculated as the percentile rank of \(d(x)\) relative to the validation distances: a low score indicates that \(x\) lies in a well-supported region, while a high score indicates that it lies farther from the training data than most observed samples. To make this assessment explicit, the algorithm also constructs a support set \(\mathsf{support}(x)\) by selecting neighbors of \(x\) within a closeness threshold \(\tau_{\text{close}}\), which identifies how many and which training configurations support the suggestion. By default, \sysname uses \(k=20\) and sets \(\tau_{\text{close}}\) to the \(5^{\text{th}}\) percentile of the validation distance distribution; both can be adjusted by the user.

\subsubsection{Model Uncertainty (UQ)}
\label{sec:uq}
However, data support alone does not establish trustworthiness. In HPC datasets, nearby configurations often have similar objective values, so a configuration can lie in a data-supported region and still have an unstable prediction if plausible models disagree there. This means that the local input--performance mapping is not well determined. To capture this second trustworthiness signal, Algorithm~\ref{alg:assess_reliability} adds a second component based on ensemble disagreement. It uses an ensemble of \(P\) independently trained models to compute prediction variance for each validation sample \(v_i \in X_{\text{val}}\) and for the generated configuration \(x\) (Lines~13--14). These models are separate from the surrogate \(f_{\theta}\) used for candidate generation. They are trained on perturbed versions of \(X_{\text{train}}\). The resulting uncertainty for \(x\), denoted \(u(x)\), is then normalized as \(UQ(x)\). Specifically, \(UQ(x)\) is defined as the percentile rank of \(u(x)\) relative to the validation uncertainties \(\{u(v_i)\}\) (Line~15). Low values indicate typical prediction stability. High values indicate that the prediction is sensitive to model variation. Following prior work, by default \sysname uses \(P=5\) ensemble members to balance uncertainty fidelity and training cost~\cite{lakshminarayanan2017simplescalablepredictiveuncertainty}.
\subsection{Actionable Outcome}
\label{sec:actionable-outcome}
\begin{figure}[t]
    \centering
    \includegraphics[width=\linewidth]{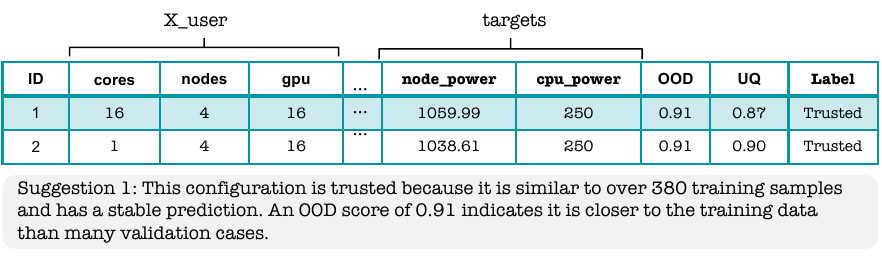}
    \caption{Illustration of how \sysname displays and explains the suggested configurations.}
    \label{fig:actionable-outcome}
\end{figure}
\begin{figure}[t]
    \centering
    \includegraphics[width=\linewidth]{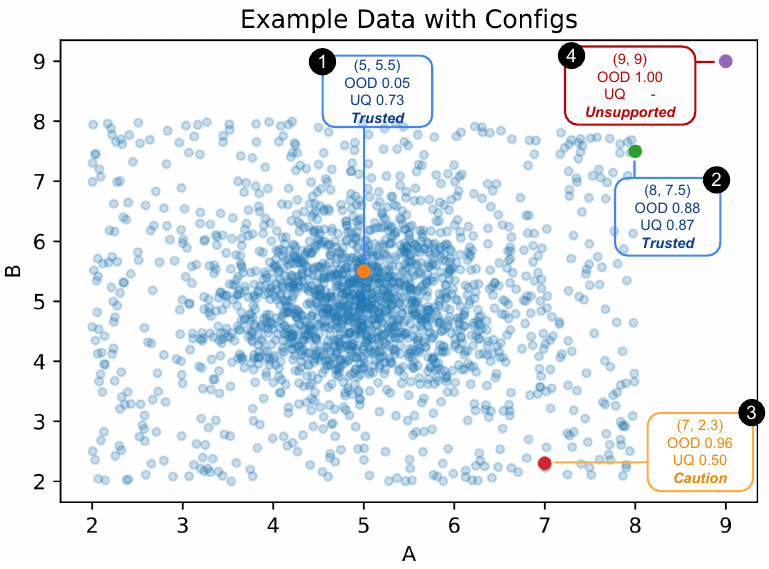}
    \caption{Illustrative Example: Observed data points and generated configurations with respective (A, B) values, OOD and UQ scores, and label.}
    \label{fig:toy-data}
\end{figure} 
Figure~\ref{fig:actionable-outcome} shows how \sysname presents trustworthiness in a real-data example. For each query, \sysname returns the top-\(\gamma\) configurations together with predicted targets, trustworthiness labels, OOD scores, and UQ scores. In this example, the top-ranked configuration is labeled \textit{trusted} because its OOD and UQ scores are below the unreliability thresholds and it has more than 380 supporting samples. Each returned configuration also includes a short natural-language explanation of its label. 
To illustrate the remaining labels, Figure~\ref{fig:toy-data} shows a toy example with configuration parameters \(A\), \(B\), and target \(C\). There, Configuration~(1) is labeled \textit{trusted} with about 278 supporting samples, whereas Configuration~(3) is labeled \textit{caution} with about 14 supporting samples, which triggers a suggestion for the user to collect additional data in that region.

\section{Scalable Implementation}
\label{sec:implementation}

\subsection{Subset Sampling.}
For large datasets such as used in this study ($~475M$ samples) When surrogate construction is required, \sysname applies subset sampling to reduce training cost while preserving representativeness. Following the best practice in the literature, we implement an uncertainty-guided loss-proportional subsampling~\cite{pmlr-v28-mineiro13} method that has been shown to be effective, specifically, for datasets with a highly variable mixture of data. Moreover, datasets ove

\subsection{Data Preprocessing}
\label{sec:data-preprocessing}
Before generation, \sysname preprocesses the uploaded dataset by: (1) removing samples with missing values in required features; (2) dropping user-specified columns; (3) applying subset sampling to reduce data if data size is large (a toggle option on the front-end selected by the user); 
(4) then splitting the resulting dataset into 80\% training and 20\% validation partitions. Numerical features are normalized using statistics from \(\mathcal{X}_{\text{train}}\), categorical features are label encoded, and trustworthiness scores are computed relative to \(\mathcal{X}_{\text{val}}\). For \modify and \whatif queries, the user selects a baseline configuration from \(\mathcal{X}_{\text{train}} \cup \mathcal{X}_{\text{val}}\). For \reco queries, no user-provided baseline is required; instead, \sysname first sub-selects samples by applying the user-provided constraints as filters, and then randomly chooses a baseline from the resulting set. If no sample satisfies all user constraints, \sysname initializes the search from an observed configuration that satisfies the largest subset of constraints, so generation still starts from a data-supported point.
\subsection{Hybrid Parallelization for Scalability}
\label{sec:parallel}
To scale to large datasets, in addition to data reduction through subset sampling, we implement \sysname using a hybrid parallel execution model that combines Dask, MPI, and multiprocessing. After the dataset is sharded across MPI ranks, each rank operates on local partitions to preserve data locality. Preprocessing and filtering stages are perfectly parallel. Candidate
generation is performed independently by multiprocessing workers within each
rank using distinct random seeds, enabling diversity without coordination.
After local generation, Rank~0 aggregates valid candidates, removes duplicates,
and computes trustworthiness labels using Algorithm~\ref{alg:assess_reliability}.
This design limits communication to a single aggregation phase and efficiently
exploits both inter- and intra-node parallelism.

\subsection{Interactive Querying Through the Front-End}
\label{sec:chatbot}
\begin{figure}[t]
    \centering
    \includegraphics[width=\linewidth]{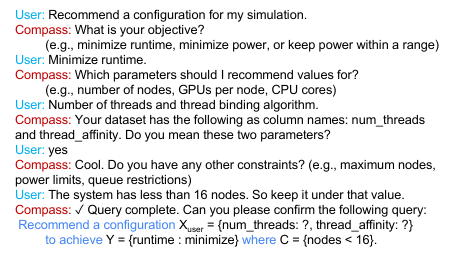}
    \caption{Example user–\sysname interaction showing how natural-language intent is refined through clarification into a structured \reco query; the same interaction pattern applies to \modify and \whatif queries. The larger view of the GUI is presented in Figure~\ref{fig:compass-ui} in the Appendix.}
    \label{fig:reco-query-example}
\end{figure}
\sysname includes a query assistant because the generation engine requires a structured specification, whereas users express their goals in natural language. To bridge this gap, the ReactJS-based chat bot (Figures~\ref{fig:reco-query-example} and~\ref{fig:compass-ui} in Appendix) asks targeted follow-up questions and incrementally maps the user's answers to a query template. 


\section{Experimental Setup}
\label{sec:setup}
\begin{table}[t]
\tiny
\caption{
Dataset characteristics and sample counts after subset sampling, showing original \#samples $\rightarrow$ sub-sampled \#samples (\% reduction); and the dataset size in MB. datasets without arrows are small and therefore not sub-sampled. \textcolor{blue}{HotCloudPerf} is a classification dataset. The MIT Supercloud dataset has $1.3$ billion samples with $126GB$ size and has been included to test the scalability of \sysname.
}
\label{tab:subset-results}
\resizebox{\columnwidth}{!}{%
\begin{tabular}{|p{1.5cm}|p{1.5cm}|p{3cm}|}
\hline
\textbf{Dataset \& \#-Samples} & \textbf{Representation} & \textbf{Features \& Target} \\
\hline

Monet \cite{jha2019monet} $800000 \rightarrow 180000$ (77\%), 112.38MB & Network-level Telemetry & coord\_X, coord\_Z, coord\_Y, X+\_USED\_BW,
\textbf{X+\_INQ\_STALL, X+\_CREDIT\_STALL}\\
\hline

PM-100 \cite{antici2023pm100} $231238 \rightarrow 41623$ (82\%), 8.55 MB & Scheduler-level Power-Performance Traces & cores\_per\_task, job\_state, num\_cores\_req, num\_nodes\_req, num\_tasks, time\_limit, num\_gpus\_req, mem\_req,
{\textbf{node\_power, mem\_power, cpu\_power, run\_time}} \\
\hline

BUTTER-E \cite{tripp2024measuringenergyconsumptionefficiency} $29644$, 15.70MB & Application-level Tuning & size, depth, learning\_rate, batch\_size, optimizer, is\_gpu, num\_reps, {\textbf{run\_time, energy, power, std\_power, std\_energy}} \\
\hline

CoMD \cite{comd} $553$, 0.03 MB & Application Execution Profile & application, algorithm, bw\_level, task\_count, thread\_count, perf\_variation, {\textbf{runtime, power\_cap}} \\
\hline

FT \cite{NPBMZ}~$554$, 0.03 MB& Application Execution Profile & application, algorithm, bw\_level, task\_count, thread\_count, perf\_variation, {\textbf{runtime, power\_cap}} \\
\hline

\textcolor{blue}{HotCloudPerf} \cite{samsi2021mitsuperclouddataset} $100000 \rightarrow 18000$ (82\%), 4.32MB & Cloud Workload Traces & NCPUS, AllocCPUS, ReqCPUS, is\_gpu, ElapsedRaw, CPUTimeRAW, NNode, {\textbf{State}} \\
\hline

NPB-MZ~\cite{NPBMZ} (Benchmark Classes: A, B, C) $675$& Hardware Performance Counters & application\_name,num\_nodes,num\_processes,thr\_aff,prob\_cls,cpu\_time\_i,papi\_br\_e,papi\_br\_i, papi\_tot\_e,papi\_tot\_i,{\textbf{cpu\_time\_e}} \\
\hline
\textbf{MIT Supercloud \cite{chu2023hotcloudperf} 
$\sim\!1.3B\!\rightarrow\!\sim\!258M$ (95\%), 126GB} 
& HPC Production Facility-scale Monitoring & NCPUS, AllocCPUS, ReqCPUS, is\_gpu, ElapsedRaw, CPUTimeRAW, NNode, {\textbf{State}} \\
\hline

\end{tabular}}
\end{table}
\subsection{Datasets}

Table~\ref{tab:subset-results} summarizes the \datasetcount datasets used in our experiments, including sample counts before and after subset sampling, original dataset size, the representative class of dataset, key features, and target metrics. 
Taken together, these datasets span distinct layers of the computing ecosystem, including interconnect telemetry, facility-scale monitoring, job-level operational traces with time-series data, application tuning profiles, and hardware-counter-based execution profiles, across both HPC and cloud settings. The workload set includes both scientific applications and ML workloads. 

\subsection{Experimental System \& Hyperparameter Tuning}
We run experiments on two TACC systems: Lonestar6 and Frontera. On Lonestar6, we used AMD EPYC 7763 CPUs and NVIDIA A100 GPUs, whereas on Frontera, we used Intel Xeon Platinum 8280 CPUs and NVIDIA RTX 5000 GPUs. We use Optuna~\cite{akiba2019optuna} to automate hyperparameter tuning for surrogate model training. Based on our ablation study, we set \(\lambda_{\mathrm{prox}} = \lambda_{\mathrm{cons}} = \lambda_{\mathrm{div}} = 1\) in all experiments because enabling all loss components yielded the best penalized MAPE.

\subsection{Evaluation Protocol, Baselines, and Reproducibility}
\textbf{Evaluation Protocol.} Reconstruction error alone is not sufficient, because a recommendation can be close to the target and still violate domain constraints. Therefore, we compare methods using penalized MAPE, defined as the sum of a normalized reconstruction term and a normalized constraint-violation term. Here, \(\texttt{diff\_norm}\) measures relative prediction error, while \(\texttt{penalty\_norm}\) is zero when all constraints are satisfied and increases as more constraints are violated. 
Lower values indicate better reconstruction of the requested query and lower domain constraint violation.
We average this penalized error over the test set. Unless noted otherwise, we report penalized MAPE throughout.
\textbf{Baselines.} We compare \sysname against widely used baselines from the literature for counterfactual generation (DiCE), autotuner (BoFire~\cite{durholt2025bofire}), LLM-based tabular data generation (\bgreat~\cite{borisov2023language}), and TABCF~\cite{panagiotou2024tabcf}.
\textbf{Statistical Analysis.} We report 95\% confidence intervals for penalized MAPE, computed across test samples from disjoint held-out partitions.

\section{Evaluation}
\label{sec:results}
We evaluate \sysname by running the following experiments:
    (1) Surrogate model selection. 
    (2) $C^3G$ Loss Ablation Study.
    (2) Effectiveness of \reco, \modify, and \whatif. 
    (3) Overhead, scalability, and reproducibility. 
    (4) Validation.
    (5) Applicability of \sysname for job scheduling.
\subsection{Surrogate Model Selection}
\label{sec:result-surrogate}

\begin{table}[t]
\tiny
\centering
\caption{Evaluate the performance of different surrogate models, as explored by the \texttt{Model Selection} component of \sysname. LGBM stands for LightGBM and RF stands for Random Forest. The \(\times\) for SVR on Monet indicates that SVR did not converge within 5 hours, even on the 20\% subset-sampled data. 
}
\label{tab:all_models_results}
\resizebox{\columnwidth}{!}{
\begin{tabular}{|p{1.0cm}|
c|c|c|
c|c|c|}
\hline
\textbf{Dataset} &
\textbf{XGBoost} & 
\textbf{LGBM} & 
\textbf{RF} & 
\textbf{SVR} & 
\textbf{MLP} & 
\textbf{Ridge} \\ \hline

Monet & 0.98 & 0.76 & \tablecellhl{0.75} & $\times$
& 1.1 & 5.2e4 \\ \hline
PM-100 & 0.21 & 0.04 & \tablecellhl{0.02} & 0.04 & 0.12 & 0.05 \\ \hline
BUTTER-E & 0.02 & 0.01 & \tablecellhl{0.01} & 0.04 & 0.02 & 0.03 \\ \hline
CoMD & 0.05 & 0.05 & \tablecellhl{0.05} & 0.07 & 0.17 & 0.04 \\ \hline
FT & 0.07 & 0.08 & \tablecellhl{0.06} & 0.14 & 0.11 & 0.05 \\ \hline
\end{tabular}
}
\end{table}



The objective of this experiment is to determine which surrogate model should be used when the user does not provide one. Table~\ref{tab:all_models_results} answers this question by comparing surrogate models across datasets, where the rows are datasets, the columns are model families, and the reported value is MAPE. We evaluate XGBoost~\cite{chen2015xgboost}, LightGBM~\cite{ke2017lightgbm}, Random Forest~\cite{biau2016random}, Support Vector Regression (SVR)~\cite{awad2015support}, Multilayer Perceptron (MLP)~\cite{rumelhart1986learning}, and Ridge~\cite{mcdonald2009ridge}. For each dataset, we split the data into 80\% training and 20\% validation, compute validation MAPE using 5-fold cross-validation on the training split, and select the model with the lowest loss.

\textbf{Observations.} Table~\ref{tab:all_models_results} shows that Random Forest achieves the lowest MAPE on most datasets, so we use it as the default surrogate in the remaining experiments. 
The \(\times\) for SVR on Monet indicates that SVR did not finish training even after 5 hours, despite using a 20\% subset-sampled data. This failure is caused by both the number of features and the feature-space complexity: Monet contains many network-counter-derived features, and SVR scales poorly in such settings.
This result motivates training a dataset-specific surrogate. Currently, \sysname trains surrogate models from scratch. In the future, supporting a more efficient adaptation method such as transfer learning or fine-tuning may improve the performance of \sysname further, and is an orthogonal issue.

\subsection{$C^3G$ Loss Ablation Study}
\label{sec:ablation-study}

Table~\ref{tab:ablation} answers this question on the CoMD dataset, where the first three columns show the loss components and the remaining columns report the average penalized MAPE (PM), OOD, UQ, 95\% confidence interval, and constraint violations for the three query types. We keep the validity term \(\mathcal{L}_{\mathrm{valid}}\) active in all runs because it is the core term that enables the generative model to learn. We then isolate the effect of the remaining three terms in Equation~\ref{eqn:c3g_cf_loss} 
by toggling \(\lambda_{\mathrm{prox}}, \lambda_{\mathrm{div}}, \lambda_{\mathrm{cons}} \in \{0,1\}\). 
We evaluate each loss configuration on 10 test requests for each of the three query types, for a total of 240 runs.

\textbf{Observations.} Table~\ref{tab:ablation} shows that using all three loss terms gives the lowest penalized MAPE for all three query types and reduces domain-constraint violations to \(0\%\). In contrast, removing proximity causes the largest increase in violations for \reco and \whatif, because both are counterfactual-generation queries: they must satisfy the target while changing a baseline configuration as little as possible. One of our novel contribution to the $C^3G$ loss---the domain-constraint component---is particularly important for \modify. Without it, the violation rate rises to \(44.4\%\), showing that the model can generate configurations that appear plausible but are not feasible to run in practice. The diversity term has less effect on penalized MAPE than proximity and constraints, and further investigation shows that diversity weight starts to matter when user chooses to explore a large number of configuration space. We test this hypothesis by asking \sysname to generate 100 configurations for a single query with a large value of $\lambda_{div} = 90$. This scenario indicates that a user prefers to explore the configuration space further from the data they have collected. 

\subsection{Effectiveness of \reco}
\label{sec:results-compass-vs-others}

\begin{table}[t]
\tiny
\centering
\caption{
All errors are weighted MAPE (lower is better). Blue cells denote best. Grey indicates the method ran out of memory, exceeded 12 hours of training allocation time or infer a single suggestion, or does not perform that operation.}
\label{tab:be-great-comp-summary}
\resizebox{\columnwidth}{!}{
\begin{tabular}{|l|p{1cm}|p{1cm}|p{1cm}|p{1cm}|p{1cm}|}
\hline
\textbf{Dataset} & 
{\textbf{\sysname}} & {\textbf{\begreat}} & {\textbf{\bofire}}
&{\textbf{DiCE}} &
{\textbf{TABCF}}\\
\hline

\multicolumn{6}{|c|}{\tablecellbl{\textbf{\reco}}} \\ 
\hline
Monet  &  {$0.11\!\pm\!0.04$}  & {\cellcolor{gray!20}}   & \tablecellhl{$0.09\!\pm\!0.06$} &  \multicolumn{2}{c|}{\cellcolor{gray!20}{}} \\ \cline{1-4}
PM-100 &  \tablecellhl{$0.09\!\pm\!0.03$} & \cellcolor{gray!20} & \cellcolor{gray!20} &\multicolumn{2}{c|}{\cellcolor{gray!20}{Do not}}\\ \cline{1-4}
BUTTER-E &  \tablecellhl{$0.01\!\pm\!0.01$}& {$6.62\!\pm\!0.00$}  & {$1.04\!\pm\!0.20$} & \multicolumn{2}{c|}{\cellcolor{gray!20}{perform}} \\ \cline{1-4}
CoMD   & \tablecellhl{$0.19\!\pm\!0.11$}  &   {\cellcolor{gray!20}} & {$0.44\!\pm\!0.19$} & \multicolumn{2}{c|}{\cellcolor{gray!20}{\reco.}}\\ \cline{1-4}
FT    &  \tablecellhl{$0.12\!\pm\!0.08$} & {\cellcolor{gray!20}{}} &  {$0.37\!\pm\!0.13$} & \multicolumn{2}{c|}{\cellcolor{gray!20}{}} \\ \cline{1-4}
NPB-MZ  &  {$0.17\!\pm\!0.11$} & {\cellcolor{gray!20}{}} & \tablecellhl{$0.09\!\pm\!0.09$} & \multicolumn{2}{c|}{\cellcolor{gray!20}{}} \\ \cline{1-4}
\textcolor{blue}{HotPerfCloud}  & \tablecellhl{$2.4\mathrm{e}{-3}\!\pm\!0.00$}  &  \cellcolor{gray!20} & {\cellcolor{gray!20}} &\multicolumn{2}{c|}{\cellcolor{gray!20}{}}\\ \hline

\multicolumn{6}{|c|}{\tablecellgr{\textbf{\modify}}} \\ \hline
Monet   & \tablecellhl{$0.11\!\pm\!0.05$} &  \multicolumn{2}{c|} {\cellcolor{gray!20}} & {\cellcolor{gray!20}} & {$0.26\!\pm\!0.13$} \\ \cline{1-2}\cline{5-6}
PM-100  &\tablecellhl{$0.13\!\pm\!0.04$} & \multicolumn{2}{c|} {\cellcolor{gray!20}}  & {$0.12\!\pm\!0.07$} & {$0.32\!\pm\!0.09$} \\ \cline{1-2}\cline{5-6}
BUTTER-E & \tablecellhl{$0.03\!\pm\!0.05$} & \multicolumn{2}{c|} {\cellcolor{gray!20}{Unable}} & {$0.41\!\pm\!0.32$} & {$0.13\!\pm\!0.07$} \\\cline{1-2}\cline{5-6}
CoMD   & \tablecellhl{$0.13\!\pm\!0.16$} & \multicolumn{2}{c|} {\cellcolor{gray!20}{to}} & {$0.15\!\pm\!0.23$} & {$0.73\!\pm\!0.28$} \\ \cline{1-2}\cline{5-6}
FT    & \tablecellhl{$0.11\!\pm\!0.06$}  & \multicolumn{2}{c|} {\cellcolor{gray!20}{generate.}} & {$0.13\!\pm\!0.06$} & {$0.85\!\pm\!0.13$} \\ \cline{1-2}\cline{5-6}
NPB-MZ & {$0.23\!\pm\!0.00$}  & \multicolumn{2}{c|} {\cellcolor{gray!20}} & \tablecellhl{$0.20\!\pm\!0.04$} & {$0.16\!\pm\!0.12$} \\ \cline{1-2}\cline{5-6}
\textcolor{blue}{HotPerfCloud}  & \tablecellhl{$0.0\!\pm\!0.0$} & \multicolumn{2}{c|}{\cellcolor{gray!20}}  & {$0.0\!\pm\!0.0$} & {\cellcolor{gray!20}} \\ \hline

\multicolumn{6}{|c|}{\tablecellor{\textbf{\whatif}}} \\ \hline

Monet &  \cellcolor{gray!20} &\multicolumn{2}{c|} {\cellcolor{gray!20}}  & \cellcolor{gray!20} & \cellcolor{gray!20} \\ \cline{1-2}\cline{5-5}
PM-100 & \tablecellhl{$5.94\mathrm{e}{-3} \!\pm\! 5.49\mathrm{e}{-3}$} & 
\multicolumn{2}{c|} {\cellcolor{gray!20}{Unable}} & {$0.01\!\pm\!0.02$} & \cellcolor{gray!20} Unable \\ \cline{1-2}\cline{5-5} 
BUTTER-E & \tablecellhl{$0.0\!\pm\!0.0$} & \multicolumn{2}{c|}{\cellcolor{gray!20}{to}} & {$0.40\!\pm\!0.32$} & \cellcolor{gray!20} to  \\ \cline{1-2}\cline{5-5}
CoMD & \tablecellhl{$0.03\!\pm\!0.03$} & \multicolumn{2}{c|}{\cellcolor{gray!20}{generate.}} & {$0.05\!\pm\!0.04$} & \cellcolor{gray!20} generate. \\ \cline{1-2}\cline{5-5}
FT & \tablecellhl{$0.10\!\pm\!0.02$} & \multicolumn{2}{c|}{\cellcolor{gray!20}} & \tablecellhl{$0.10\!\pm\!0.02$} & \cellcolor{gray!20} \\ \cline{1-2}\cline{5-5}
NPB-MZ & {$0.19\!\pm\!0.08$} & \multicolumn{2}{c|}{\cellcolor{gray!20}} & \tablecellhl{$0.08 \!\pm\! 0.00$} & \cellcolor{gray!20} \\ \cline{1-2}\cline{5-5}
\textcolor{blue}{HotPerfCloud} & \tablecellhl{$1.37\mathrm{e}{-3}\!\pm\!4.47\mathrm{e}{-4}$} & \multicolumn{2}{c|}{\cellcolor{gray!20}} &  {$1.37\mathrm{e}{-3}\!\pm\!4.8\mathrm{e}{-3}$}   & \cellcolor{gray!20} \\ \hline

\end{tabular}
}
\end{table}

The objective of this experiment is to test whether our unified formulation of \reco yields more accurate recommendations than two alternatives: \begreat, a conditional tabular generator based on an LLM, and \bofire, an iterative black-box autotuner. 
\dice and \tabcf are excluded because they require a user-provided initial configuration, which \reco does not assume. 
We evaluate each method by masking part of a real held-out hardware configuration and asking it to recover the missing settings from the target \(Y\); if it can 
reconstruct that observed configuration, the reconstruction error itself serves as validation of real-system behavior.

\textbf{Observations.}
From Table~\ref{tab:be-great-comp-summary}, \sysname returns valid recommendations in all experiments. In contrast, \bofire fails to return a recommendation for the PM-100 and HotPerfCloud datasets. 
Further investigation on PM-100 shows that the query fixes three features, while 23 domain constraints must still be satisfied. This leaves a small feasible region, so \bofire must search through many invalid candidates before finding a valid one, and may fail within the search budget of this autotuner. Table~\ref{tab:appendix-domain-constraints} lists example constraints for this dataset. \begreat does not produce usable recommendations for most datasets even after 16 hours of training and 1,000 epochs, supporting our claim that unconstrained generation is not suitable for configuration generation. For BUTTER-E, it generates some outputs, but the penalized MAPE shows that these poorly reconstruct the target instances. Overall, \sysname outperforms \bofire in 5 of the 7 experiments reported in Table~\ref{tab:be-great-comp-summary}. 

\subsection{Effectiveness of \modify}
\label{sec:modify}
The objective of this experiment is to evaluate the effectiveness of \sysname's $C^3G$ loss function compared to other counterfactual generation libraries such as \dice and \tabcf~\cite{panagiotou2024tabcf} for the \modify query type. All methods operate on the same initial configurations and use an identical surrogate model $f_{\theta}$, ensuring that observed differences arise solely from their configuration-generation mechanisms.

\textbf{Observations.}
Table~\ref{tab:be-great-comp-summary} shows that overall, the penalized MAPE for \sysname is consistently lower than that of other counterfactual generation libraries. This difference arises because \sysname encodes domain constraints directly into the generative loss and optimizes a hinge-based target loss (\( \mathcal{L}_{\text{valid}} \)), whereas \dice relies on a point-based loss. As a result, \sysname generates counterfactuals that more faithfully satisfy the target while remaining feasible. The only exception is for NPB-MZ where \dice performs better, however \sysname has the lower variability which suggests that our \sysname is more reliable and less sensitive to randomness. 


\subsection{Effectiveness of \whatif}
\label{sec:whatif}
This experiment evaluates the effectiveness of \sysname in solving the forward problem.
The specific queries for each dataset are summarized in Table~\ref{tab:results-appendix}.
We compare \sysname with DiCE and TABCF.
For evaluation, we use the held-out test set (\texttt{X\_test}) to generate reference baselines by issuing \whatif queries that change a set of input features while keeping all other input features fixed, and we evaluate the resulting change in $Y_j$ by comparing the generated configurations (\texttt{X\_gen}) against the original ground truth test sample. If a method can accurately recover a held-out configuration that was actually observed on the real hardware, then it has demonstrated that it can reproduce valid real-system behavior.

\textbf{Observations.}
From Table~\ref{tab:be-great-comp-summary}, we observe that \sysname achieves low error on most \whatif queries. The reason is that what-if queries require structured local perturbations of a reference configuration, and \sysname is designed to generate such perturbations while enforcing feasibility. In contrast, some baselines are unable to generate outputs at all in this setting, which suggests that generic tabular generation is less effective when the task requires targeted changes to \(Y\) under query-specific constraints.
For NPB-MZ, \sysname has a higher error than \dice. A possible reason is that \dice explicitly promotes sparsity, which can lead to fewer changed features and therefore lower \(L_1\)-based reconstruction error. In contrast, \sysname may generate denser but still valid perturbations, which can increase the reconstruction error even when the generated configurations remain feasible.

\subsection{Validation}
\label{sec:validation}
\begin{table}[t]
\tiny
\caption{Summary of validation results.}
\label{tab:validation}
\setlength{\tabcolsep}{4pt}
\renewcommand{\arraystretch}{0.95}
\resizebox{\columnwidth}{!}{%
\begin{tabular}{|p{1.2cm}|p{4.95cm}|}
\hline

\rowcolor{bll}
\multicolumn{2}{|c|}{\rule{0pt}{1.9ex}\textbf{\reco. ML Task: Inverse Problem}\rule[-0.6ex]{0pt}{0pt}}\\ \hline

\textbf{Monet} &
\sysname suggested mesh coordinates with \texttt{nettopo\_mesh\_coord\_Z} and \texttt{nettopo\_mesh\_coord\_X} placed closer together, which matches Jha et al.~\cite{jha2019monet}'s finding that placing communicating tasks at nearby mesh coordinates reduces hop count and improves network performance.
\\ \hline

\textbf{PM-100} &
\sysname suggested configurations that satisfy the system constraint \texttt{num\_gpus\_req = 4 * num\_nodes\_req}, which matches the PM-100 platform configuration described in the dataset.
\\ \hline

\textbf{BUTTER-E} &
\sysname suggested batch size 256, the Adam optimizer, and a low-to-medium depth of 7, which matches Tripp et al.~\cite{tripp2024measuringenergyconsumptionefficiency}, who reported that low-depth models with batch size 256 and Adam reduce power consumption.
\\ \hline

\textbf{FT} &
\sysname suggested the \(pak\) algorithm, which matches Patki et al.~\cite{patki2019performance}, who reported that \(pak\) yields the least run-to-run performance variation.
\\ \hline

\rowcolor{sgl}
\multicolumn{2}{|c|}{\rule{0pt}{1.9ex}\textbf{\modify. ML Task: Counterfactual Reasoning.}\rule[-0.6ex]{0pt}{0pt}}\\ \hline

\textbf{Monet} &
\sysname suggested reducing \texttt{nettopo\_mesh\_coord\_Y} by 7.14\%, which matches Jha et al.~\cite{jha2019monet}'s observation that reducing Z credit stalls requires redistributing the network layout.
\\ \hline

\textbf{PM-100} & 
\sysname suggested reducing CPU cores by 68\% to meet the target, which matches the known overprovisioning behavior in PM-100 configurations.
\\ \hline

\textbf{BUTTER-E} &
\sysname suggested a moderate-to-high depth of 15 rather than the maximum depth, which matches Tripp et al.~\cite{tripp2024measuringenergyconsumptionefficiency}, who showed that deeper models do not always improve the accuracy--energy trade-off.
\\ \hline

\textbf{FT} &
\sysname suggested changing the algorithm from \texttt{rand} to \texttt{spr}, which matches Patki et al.~\cite{patki2019performance}, who showed that this change reduces performance variation by about 10\%.
\\ \hline

\rowcolor{orl}
\multicolumn{2}{|c|}{\rule{0pt}{1.9ex}\textbf{\whatif. ML Task: Forward Problem.}\rule[-0.6ex]{0pt}{0pt}}\\ \hline

\textbf{Monet} & 
\sysname predicted a 54.32\% increase in link stalls, which matches published observations on Blue Waters that lower credit availability increases network stall events~\cite{jha2019monet}.
\\ \hline

\textbf{PM-100} &
\sysname predicted that increasing the number of requested GPUs would also require increases in node and memory power consumption, which matches the coupling among these resources in PM-100 configurations.
\\ \hline

\textbf{BUTTER-E} &
\sysname predicted that doubling the depth leads to only a 3.41\% runtime increase, which matches Tripp et al.~\cite{tripp2024measuringenergyconsumptionefficiency}, who showed that increasing depth does not always produce a proportional increase in power or runtime.
\\ \hline

\textbf{FT} &
\sysname predicted that changing the algorithm from \(rand\) to \(pak\) changes run-to-run performance variation by 8.4\% and runtime by 12.1\%, which matches the trends reported by Patki et al.~\cite{patki2019performance}.
\\ \hline

\end{tabular}%
}
\end{table}
\textbf{Empirical Validation on Hardware.} We need this experiment to test whether \sysname's recommendations remain accurate when checked against real executions on the target hardware, rather than only against offline data. To do so, we matched all recommendation rows for which measured performance data were available and compared the predicted CPU time against the observed CPU time extracted from HPCToolkit. Out of 202 recommendation rows, 185 were matched to measured runs. For example, for \texttt{sp-mz} with 2 nodes, 3 processes, random affinity, and class C, \sysname predicted 272.08 seconds, while the measured CPU time was 297.32 seconds, corresponding to an error of 8.49\%. Across all 185 matched rows, the mean absolute error was 39.73 seconds and the MAPE was 13.31\%. Overall, these results show that \sysname's predictions track measured execution time with moderate error. 

\textbf{Validation against Published Results.} We validate \sysname by reproducing recommendations that agree with established findings, not just with held-out samples. To do so, for each dataset in Table~\ref{tab:validation}, we formulate one query from the corresponding paper and check whether \sysname reproduces the published observation. This form of validation is important because many of these observations were originally obtained through manual analysis, whereas \sysname recovers them automatically in a few seconds. 

\textbf{Validation using Analytical Performance Models.}
In addition, we validate \sysname against published analytical performance models through benchmarking and show that it can reproduce their expected behavior. Due to space limitations, we defer the analytical-model-based validation results to Appendix~\ref{tab:syn-mape}.
\subsection{Overhead and Reproducibility}
\label{sec:overhead}
\textbf{Overhead.}
This experiment evaluates the runtime overhead of \sysname across query types and compares it with \begreat. 
On the largest dataset MIT Supercloud~\cite{chu2023hotcloudperf} (\(\sim\)1.3B samples), \sysname completes end-to-end execution in 419.71s and generates 20 suggestions in 60.78s, while no baseline can handle this dataset. On BUTTER-E, \sysname takes 1.74s, whereas \begreat takes 140s to generate one configuration, giving \sysname up to \(80\times\) faster inference. \begreat also incurs high training cost: 16h on Monet, 4.5h on PM-100, 2.5h on BUTTER-E, 2h on CoMD, and 1.5h on both FT and NPB using a single A100 (80GB) GPU.
\textbf{Reproducibility.}
The subset selection process is deterministic: given fixed data, random seeds, and a retention fraction, the same samples are selected across runs, ensuring reproducibility.

\section{Benefit of \sysname for Job Scheduling}
\label{sec:application}
We evaluate \sysname for HPC job scheduling using the exact workload traces and discrete-event simulator released with \textsc{ModelX}~\cite{dey2025modelx,ai-scheduler}. We use the published artifact unchanged, split jobs into 80--20 train--test partitions, train \sysname on the 80\%, and evaluate on the held-out 20\%. For each test job, we issue a \modify query that changes only \texttt{node\_count} to achieve at least a 10\% runtime reduction, while fixing all other parameters through \(C\). We use \texttt{random\_forest} as the surrogate model and repeat the experiment over five independent splits. This setup enables direct comparison with user-specified, heuristic-oracle, and \textsc{ModelX} baselines under the same workload and system conditions. We observe that \sysname reduces average job turnaround time by \(65.93\%\), \(28.03\%\), and \(25.53\%\) relative to user-specified, heuristic-oracle, and \textsc{ModelX}, respectively. The reason is that \sysname recommends \(80.93\%\), \(35.6\%\), and \(67.77\%\) fewer nodes on average compared to the above mentioned baselines, which increases system availability and reduces queueing delay. These results highlight the benefit of \sysname-guided job scheduling in HPC. 

\section{Related Work}
\label{sec:related-work}
\textbf{Recommendation and Autotuning (Recommend).}
In our taxonomy, autotuners and inverse-design methods instantiate the \emph{recommend} task: given objectives, find a configuration. Prior HPC autotuning frameworks (Table~\ref{tab:comparison_autotunner}) optimize configurations to minimize or maximize target metrics. Early systems such as ytopt~\cite{wu2025ytopt} and OpenTuner~\cite{ansel2014opentuner} use Bayesian optimization and ensemble search for compiler- and application-level tuning; Active Harmony~\cite{tapus2002active} targets runtime parameters via search-based navigation. Bayesian optimization is applied to loop transformations in HYPERF~\cite{park2025hyperf}, configuration search across applications and hardware via Gaussian-process–guided tuning in GPTune~\cite{liu2021gptune}, and tabular experimental design in BoFire~\cite{durholt2025bofire}. Inverse design methods in materials science and photonics similarly solve ``given properties, find a design" as a standalone optimization problem~\cite{park2023machine,zhen2025machine,invdesign-materials}.
Across these approaches, optimization is treated as a one-shot mapping from objectives to configurations. They do not support minimal edits around a baseline configuration, counterfactual reasoning over observed performance, or explicit \emph{what-if} analysis.

\textbf{Generative Modeling and Counterfactual Reasoning (Reconfigure).}
These approaches correspond to the \emph{reconfigure} task: counterfactual reasoning that minimally edits a baseline configuration to achieve a new outcome, studied largely in isolation from recommendation and forward prediction. Standard generative models, including GANs~\cite{goodfellow2020generative} and diffusion models~\cite{kotelnikov2023tabddpm}, approximate global data distributions, whereas decision support requires local, constraint-aware, goal-directed generation; distribution-matching alone can violate feasibility. Counterfactual and recourse methods such as \dice perform minimal edits to change outcomes but assume classification settings, do not account for domain constraint checking, and do not unify forward prediction or inverse modeling.

\textbf{Forward Modeling and Performance Analysis (What-if).}
In Table~\ref{tab:query-templates}, these map to \emph{what-if} queries: forward prediction from configuration to outcome.
Different tools, such as PerfExplorer~\cite{huck2005perfexplorer}, HPCToolkit~\cite{adhianto2010hpctoolkit}, and PADDLE~\cite{thiagarajan2018paddle}, represent major advances in performance analysis. PerfExplorer provides clustering, summarization, and comparative analytics to reveal performance trends across runs.  
HPCToolkit attributes resource inefficiencies to code regions across CPUs and GPUs, supporting deep attribution for hybrid parallel programs. PADDLE extends these ideas through a machine learning platform for prediction and workflow integration, enabling analysts to extract performance patterns from complex datasets.   However, these systems are largely diagnostic—they identify issues but do not generate or recommend \emph{actionable alternatives}. Similarly, PERFGEN~\cite{banday2024perfgen} synthesizes performance data to aid model training, yet its outputs serve simulation or benchmarking purposes rather 
guiding configuration optimization. 

\section{Conclusions}
\label{sec:conclusions}
We present \sysname, a novel decision support system for HPC. 
We contribute along four axes: methodological, by introducing a unified generative modeling formulation called $C^3G$; empirical, by validating it through ground-truth reconstruction, analytical-performance-model benchmarking, and reproduction of published results; engineering, by building a modular implementation that scales to traces with up to 1.3 billion samples and 126 GB of data; and comparative performance, by showing that, relative to prior generative baselines, training is up to $100x$ faster and inference is up to $90x$ faster.
Across synthetic and real HPC datasets, \sysname achieves low error and small configuration changes, showing that counterfactual reasoning is effective for producing actionable recommendations. When confidence is low, \sysname abstains and identifies configurations that should be measured next. Future work will extend \sysname with stronger surrogate models, causal structure discovery, and tighter integration with live runtime systems.

\section{Acknowledgement}

This material is based upon work supported in part by the U.S. Department of Energy, Office of Science under Award Number DE-SC0022843. This material is based upon work supported in part by the National Science Foundation under Grant No.~2443561. The authors acknowledge the Texas Advanced Computing Center (TACC) at The University of Texas at Austin for providing computational resources that have contributed to the research results reported within this paper.

\bibliographystyle{IEEEtran}
\bibliography{tzi_bb}

\subsection{Appendix}

\appendix{\sysname GUI}
\begin{figure*}[t]
    \centering
    \includegraphics[width=\linewidth]{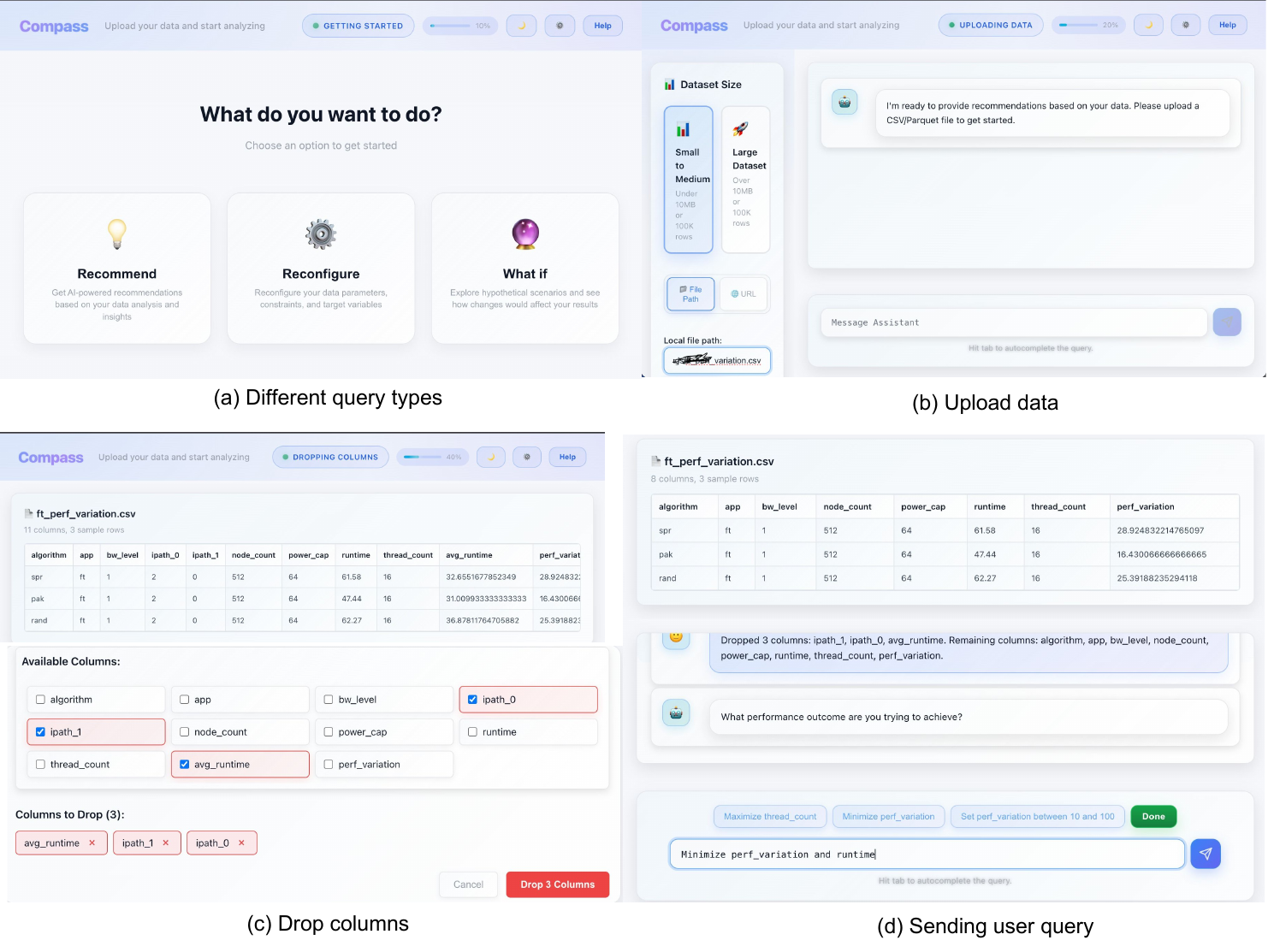}
\caption{\sysname's chat bot interface. User (a) selects a query type: \reco, \modify, or \whatif, (b) selects a dataset in \texttt{csv} or \texttt{parquet} format, (c) drops unnecessary columns, (d) responds to questions about specifying the target objectives.}
    \label{fig:compass-ui}
\end{figure*}

 \begin{table}[t]
\centering
\tiny
\caption{Different queries and \sysname's responses for all datasets.}
\label{tab:results-appendix}
\resizebox{\columnwidth}{!}{
\begin{tabular}{|p{0.06\columnwidth}|p{0.5\columnwidth}|p{0.3\columnwidth}|}
\hline
\textbf{Dataset} & \textbf{Query} & \textbf{Response} \\ \hline

\multicolumn{3}{|c|}{{\tablecellbl{\textbf{\reco}}}}\\\hline

Monet &
\cell{
\textcolor{blue}{\textbf{Recommend a configuration where}} $X = $
\texttt{\{nettopo\_mesh\_coord\_Z: ?, nettopo\_mesh\_coord\_Y: ?, nettopo\_mesh\_coord\_X: ?\}}
\textcolor{blue}{\textbf{to achieve }}
\texttt{$Y = $\{X+\_SAMPLE\_GEMINI\_LINK\_INQ\_STALL: minimized, X+\_SAMPLE\_GEMINI\_LINK\_CREDIT\_STALL: minimized, X+\_SAMPLE\_GEMINI\_LINK\_USED\_BW: minimized\}} \textcolor{blue}{\textbf{where }}\texttt{$C =$ \{\}}
}
&
\cell{
$X = $\texttt{\{nettopo\_mesh\_coord\_Z: 19.0, nettopo\_mesh\_coord\_Y: 11.0, nettopo\_mesh\_coord\_X: 22.0\}}
}
\\ \hline

PM-100 &
\textcolor{blue}{\textbf{Recommend a configuration where}}
\texttt{$X =$ \{cores\_per\_task: ?, num\_cores\_req: ?, num\_nodes\_req: ?, mem\_req: ?, time\_limit: ?\}}
\textcolor{blue}{\textbf{to achieve}}
\texttt{$Y$ = \{node\_power\_consumption: minimized, mem\_power\_consumption: minimized, cpu\_power\_consumption: minimized\}}
\textbf{\textcolor{blue}{where  }}\texttt{$C$ = \{job\_state: completed, num\_gpus\_req: 16, partition: 1\}}
&
$X = $ \texttt{\{cores\_per\_task: 1, num\_cores\_req: 64, num\_nodes\_req: 4, mem\_req: 118, time\_limit: 60\}}
\\ \hline

BUTTER-E &
\textcolor{blue}{\textbf{Recommend a configuration where}}
\texttt{$X = $ \{depth: ?, shape: ?, batch\_size: ?, optimizer: ?\}}
\textcolor{blue}{\textbf{to achieve}}
\texttt{$Y$ = \{power: minimized, runtime: minimized\}} \textcolor{blue}{\textbf{where }} \texttt{$C =$ \{is\_gpu: True, dataset: ``mnist"\}} 
&
\texttt{$X =$ \{depth: 7, shape: rectangle\_residual, batch\_size: 256, optimizer: Adam\}}
\\ \hline

CoMD &
\textcolor{blue}{\textbf{Recommend a configuration where}}
\texttt{$X = $\{bw\_level: ?, power\_cap: ?, algorithm: ?\}}
\textcolor{blue}{\textbf{to achieve}}
\texttt{$Y$ = \{perf\_variation: minimized, runtime: minimized\}} \textcolor{blue}{\textbf{where }}\texttt{$C = $\{app: CoMD\}}
&
\texttt{$X =$ \{bw\_level: 1, power\_cap: 64, algorithm: ``spr"\}}
\\ \hline

FT &
\textcolor{blue}{\textbf{Recommend a configuration where}}
\texttt{$X = $\{bw\_level: ?, power\_cap: ?, algorithm: ?, thread\_count: ?\}}
\textcolor{blue}{\textbf{to achieve}}
\texttt{$Y =$ \{perf\_variation: minimized, runtime: minimized\}} \textcolor{blue}{\textbf{where }} \texttt{$C =$ \{app: FT\}}
&
\texttt{$X =$ \{bw\_level: 4, power\_cap: 80, algorithm: ``pak", thread\_count: 16\}}
\\ \hline

\multicolumn{3}{|c|}{{\tablecellgr{\textbf{\modify}}}}\\\hline

Monet &
\textcolor{blue}{\textbf{Reconfigure}}
\texttt{$X =$ \{X+\_*\_CREDIT\_STALL: x→2x, X+\_*\_INQ\_STALL: x $\rightarrow$ 2x\}}
\textcolor{blue}{\textbf{to achieve}}
\texttt{$Y =$ \{Z+\_*\_CREDIT\_STALL: reduce by 20\%\}}
\textcolor{blue}{\textbf{where $C =$}} \texttt{\{\}}
&
\texttt{$X = $ \{Z-\_SAMPLE\_GEMINI\_LINK\_INQ\_STALL $\downarrow$93.8\%, nettopo\_mesh\_coord\_Y $\downarrow$7.14\%\}}
\\ \hline

PM-100 &
\textcolor{blue}{\textbf{Reconfigure}}
\texttt{$X$ = \{cores\_per\_task: ?, num\_cores\_req: ?, num\_nodes\_req: ?, mem\_req: ?, time\_limit: ?\}}
\textcolor{blue}{\textbf{to achieve}}
\texttt{$Y$ = \{node\_power\_consumption, mem\_power\_consumption, cpu\_power\_consumption: reduced by 10\%\}} 
\textcolor{blue}{\textbf{where}} \texttt{$C$ = \{num\_gpus\_req: 16, job\_state: completed\}}
&
\texttt{$X$ = \{cores\_per\_task $\downarrow$68\%, memory\_alloc $\downarrow$65\%\}}
\\ \hline

BUTTER-E &
\textcolor{blue}{\textbf{Reconfigure}}
\texttt{X = \{depth: ?, shape: ?, batch\_size: ?, optimizer: ?\}}
\textcolor{blue}{\textbf{to achieve}}
\texttt{Y = \{power: 10\% reduction, runtime: 10\% reduction\}}
\textcolor{blue}{\textbf{where }}
\texttt{$C$ = \{is\_gpu: True, dataset:``mnist"\}}
&
\texttt{$X =$ \{depth: 15, shape: ``trapezoid", batch\_size: 256, optimizer: ``Adam"\}}
\\ \hline

CoMD &
\textcolor{blue}{\textbf{Reconfigure}}
\texttt{$X$ = \{Bw\_level: ?, power\_cap: ?\}}
\textcolor{blue}{\textbf{to achieve}}
\texttt{$Y$ = \{perf\_variation: reduce by 10\%\}}
\textcolor{blue}{\textbf{where}}
\texttt{$C$ = \{app: CoMD\}}
&
\texttt{$X$ =
\{Bw\_level:1$\rightarrow$3,
power\_cap:\!64\!$\rightarrow$\!80\}}
\\ \hline

FT &
\textcolor{blue}{\textbf{Reconfigure}}
\texttt{$X$ = \{algorithm: ?\}}
\textcolor{blue}{\textbf{to achieve}}
\texttt{$Y$ = \{perf\_variation: reduce by 10\%\}}
\textcolor{blue}{\textbf{where}}
\texttt{$C$ = \{app: FT \}}
&
\texttt{$X$ = \{algorithm: rand→spr\}}
\\ \hline

\multicolumn{3}{|c|}{{\tablecellor{\textbf{\whatif}}}}\\\hline

Monet &
\textcolor{blue}{\textbf{What-if we change}}
\texttt{$X$ = \{Y+\_SAMPLE\_GEMINI\_LINK\_CREDIT\_STALL\!= x $\rightarrow$ x/2 , nettopo\_mesh\_coord\_X\!=\!?, Y\!=\!?, Z\!=\!?\}}
\textcolor{blue}{\textbf{how would that affect}}
\texttt{$Y$ = \{link stalls\}}
\textcolor{blue}{\textbf{where}}
\texttt{$C$ = \{\}}?
&
\texttt{$Y$ = \{Y+\_link\_stall: 54.32\% $\uparrow$ \}}
\\ \hline

PM-100 &
\textcolor{blue}{\textbf{What if we change}}
\texttt{$X$ = \{num\_gpu\_req: old$\rightarrow$2 $\times$ old\}}
\textcolor{blue}{\textbf{how would that affect}}
\texttt{$Y$ = \{node, mem, cpu power consumption\}}
\textcolor{blue}{\textbf{where}}
\texttt{$C$ = \{job\_state: completed\}}?
&
\texttt{$Y$ = \{node\_power\_consumption $\uparrow$5\%, mem\_power\_consumption $\uparrow$5\%\}}
\\ \hline

BUTTER-E &
\textcolor{blue}{\textbf{What if we change}}
\texttt{$X$ = \{depth: old$\rightarrow$2×old\}}
\textcolor{blue}{\textbf{how would that affect}}
\texttt{$Y$ = \{power, runtime\}}
\textcolor{blue}{\textbf{where}}
\texttt{$C$ = \{is\_gpu: True, dataset: ``adult"\}}?
&
\texttt{$Y$ = \{power: 0\% $\uparrow$, runtime: 3.41\% $\uparrow$\}}
\\ \hline

CoMD &
\textcolor{blue}{\textbf{What-if we change}}
\texttt{$X$ = \{algorithm: rand→pak\}}
\textcolor{blue}{\textbf{how will it affect}}
\texttt{$Y$ = \{perf\_variation\}}
\textcolor{blue}{\textbf{where}}
\texttt{$C$ = \{app: CoMD\}}?
&
\texttt{$Y$ = \{perf\_variation: $\downarrow$92.13\%\}}
\\ \hline

FT &
\textcolor{blue}{\textbf{What-if we change}}
\texttt{$X$ = \{algorithm: rand $\rightarrow$ pak\}}
\textcolor{blue}{\textbf{how will it affect}}
\texttt{$Y$ = \{perf\_variation, run\_time\}}
\textcolor{blue}{\textbf{where}}
\texttt{$C$ = \{app: FT\}}?
&
\texttt{$Y$ = \{perf\_variation 8.4\%$\downarrow$, run\_time $\downarrow$12.1\%\}}
\\ \hline

\end{tabular}
}
\end{table}
\begin{table}[t]
\tiny
\centering
\caption{Ablation of $C^3G$ loss components across query types. Best results in Blue.
For OOD and UQ, $[0, 0.95)$ = trusted, $[0.95, 0.99)$ = caution, ${\geq}0.99$ = unsupported.}
\label{tab:ablation}
\resizebox{\columnwidth}{!}{%
\begin{tabular}{|c|c|c|c|c|c|c|}
\hline
$\lambda_{\text{prox}}$ &
$\lambda_{\text{div}}$ &
$\lambda_{\mathrm{cons}}$ &
\textbf{PM} $\downarrow$ &
\textbf{OOD} $\downarrow$ &
\textbf{UQ} $\downarrow$ &
\textbf{Violations} $\downarrow$ \\ \hline

\rowcolor{bll}
\multicolumn{7}{|c|}{\textbf{\reco{}}} \\ \hline

$\times$ & $\times$ & $\times$
  & $0.31\!\pm\!0.14$ & $0.53\!\pm\!0.04$ & $0.78\!\pm\!0.02$ & $20\%$ \\ \hline
\checkmark & $\times$ & $\times$
  & $0.15\!\pm\!0.07$ & $0.46\!\pm\!0.04$ & $0.58\!\pm\!0.04$ & $0\%$ \\ \hline
$\times$ & \checkmark & $\times$
  & $0.31\!\pm\!0.14$ & $0.53\!\pm\!0.04$ & $0.78\!\pm\!0.02$ & $20\%$ \\ \hline
$\times$ & $\times$ & \checkmark
  & $0.17\!\pm\!0.05$ & $0.50\!\pm\!0.03$ & $0.72\!\pm\!0.04$ & $0\%$ \\ \hline
$\times$ & \checkmark & \checkmark
  & $0.17\!\pm\!0.05$ & $0.50\!\pm\!0.03$ & $0.72\!\pm\!0.04$ & $0\%$ \\ \hline
\checkmark & $\times$ & \checkmark
  & $0.14\!\pm\!0.07$ & $0.49\!\pm\!0.04$ & $0.57\!\pm\!0.03$ & $0\%$ \\ \hline
\checkmark & \checkmark & $\times$
  & $0.15\!\pm\!0.07$ & $0.46\!\pm\!0.04$ & $0.58\!\pm\!0.04$ & $0\%$ \\ \hline
\checkmark & \checkmark & \checkmark
  & \tablecellhl{$\mathbf{0.14\!\pm\!0.07}$} 
  & \tablecellhl{$\mathbf{0.44\!\pm\!0.04}$} 
  & \tablecellhl{$\mathbf{0.57\!\pm\!0.03}$} 
  & $\mathbf{0\%}$ \\ \hline

\rowcolor{bll}
\multicolumn{7}{|c|}{\textbf{\modify{}}} \\ \hline

$\times$ & $\times$ & $\times$
  & $0.66\!\pm\!0.13$ & $0.33\!\pm\!0.04$ & $0.82\!\pm\!0.02$ & $44.4\%$ \\ \hline
\checkmark & $\times$ & $\times$
  & $0.46\!\pm\!0.22$ & $0.46\!\pm\!0.04$ & $0.56\!\pm\!0.05$ & $33.3\%$ \\ \hline
$\times$ & \checkmark & $\times$
  & $0.66\!\pm\!0.13$ & $0.33\!\pm\!0.04$ & $0.82\!\pm\!0.02$ & $44.4\%$ \\ \hline
$\times$ & $\times$ & \checkmark
  & $0.22\!\pm\!0.08$ & $0.30\!\pm\!0.05$ & $0.69\!\pm\!0.02$ & $0\%$ \\ \hline
$\times$ & \checkmark & \checkmark
  & $0.22\!\pm\!0.08$ & $0.31\!\pm\!0.05$ & $0.69\!\pm\!0.02$ & $0\%$ \\ \hline
\checkmark & $\times$ & \checkmark
  & $0.46\!\pm\!0.22$ & $0.46\!\pm\!0.04$ & $0.56\!\pm\!0.05$ & $33.3\%$ \\ \hline
\checkmark & \checkmark & $\times$
  & $0.06\!\pm\!0.05$ & $0.43\!\pm\!0.06$ & $0.49\!\pm\!0.03$ & $0\%$ \\ \hline
\checkmark & \checkmark & \checkmark
  & \tablecellhl{$\mathbf{0.06\!\pm\!0.05}$} 
  & \tablecellhl{$\mathbf{0.43\!\pm\!0.03}$} 
  & \tablecellhl{$\mathbf{0.49\!\pm\!0.03}$} 
  & $\mathbf{0\%}$ \\ \hline

\rowcolor{orl}
\multicolumn{7}{|c|}{\textbf{\whatif{}}} \\ \hline

$\times$ & $\times$ & $\times$
  & $0.57\!\pm\!0.08$ & $0.26\!\pm\!0.03$ & $0.82\!\pm\!0.005$ & $40\%$ \\ \hline
\checkmark & $\times$ & $\times$
  & $0.075\!\pm\!0.02$ & $0.45\!\pm\!0.03$ & $0.45\!\pm\!0.02$ & $0\%$ \\ \hline
$\times$ & \checkmark & $\times$
  & $0.57\!\pm\!0.08$ & $0.26\!\pm\!0.03$ & $0.824\!\pm\!0.005$ & $40\%$ \\ \hline
$\times$ & $\times$ & \checkmark
  & $0.13\!\pm\!0.05$ & $0.22\!\pm\!0.03$ & $0.68\!\pm\!0.01$ & $30\%$ \\ \hline
$\times$ & \checkmark & \checkmark
  & $0.13\!\pm\!0.05$ & $0.22\!\pm\!0.01$ & $0.68\!\pm\!0.01$ & $0\%$ \\ \hline
\checkmark & $\times$ & \checkmark
  & $0.07\!\pm\!0.02$ & $0.43\!\pm\!0.03$ & $0.44\!\pm\!0.01$ & $0\%$ \\ \hline
\checkmark & \checkmark & $\times$
  & $0.07\!\pm\!0.02$ & $0.45\!\pm\!0.03$ & $0.45\!\pm\!0.02$ & $0\%$ \\ \hline
\checkmark & \checkmark & \checkmark
  & \tablecellhl{$\mathbf{0.07\!\pm\!0.02}$} 
  & \tablecellhl{$\mathbf{0.43\!\pm\!0.03}$} 
  & \tablecellhl{$\mathbf{0.44\!\pm\!0.01}$} 
  & $\mathbf{0\%}$ \\ \hline

\end{tabular}%
}
\end{table}
\begin{table}[t]
\centering
\tiny
\caption{Domain Constraints For PM-100 dataset}
\label{tab:appendix-domain-constraints}
\begin{tabular}{ll}
\toprule
\textbf{Category} & \textbf{Constraints} \\
\midrule

\textbf{Node} &
\begin{tabular}[t]{@{}l@{}}
$\text{num\_nodes\_req} \ge 1$ \\
$\text{num\_nodes\_alloc} \ge 0$ \\
$\text{num\_nodes\_alloc} = \text{num\_nodes\_req}$
\end{tabular} \\[0.8em]

\textbf{Core} &
\begin{tabular}[t]{@{}l@{}}
$\text{num\_cores\_req} \ge 1$ \\
$\text{num\_cores\_req} \le 32 \cdot \text{num\_nodes\_req}$ \\
$\text{num\_cores\_alloc} \ge 0$ \\
$\text{num\_cores\_alloc} \le \text{num\_cores\_req}$ \\
$\text{cores\_per\_task} \ge 1$ \\
$\text{cores\_per\_task} \le \text{num\_cores\_alloc}$
\end{tabular} \\[0.8em]

\textbf{GPU} &
\begin{tabular}[t]{@{}l@{}}
$\text{num\_gpus\_req} \ge 0$ \\
$\text{num\_gpus\_req} \le 4 \cdot \text{num\_nodes\_req}$ \\
$\text{num\_gpus\_alloc} \ge 0$ \\
$\text{num\_gpus\_alloc} = \text{num\_gpus\_req}$
\end{tabular} \\[0.8em]

\textbf{Memory} &
\begin{tabular}[t]{@{}l@{}}
$\text{mem\_req} \ge 0$ \\
$\text{mem\_req} \le 256 \cdot \text{num\_nodes\_req}$ \\
$\text{mem\_alloc} \ge 0$ \\
$\text{mem\_alloc} = \text{mem\_req}$
\end{tabular} \\[0.8em]

\textbf{Time} &
\begin{tabular}[t]{@{}l@{}}
$\text{run\_time} \ge 0$ \\
$\text{time\_limit} \ge 0$ \\
$\text{run\_time} \le \text{time\_limit}$
\end{tabular} \\[0.8em]

\textbf{Task} &
\begin{tabular}[t]{@{}l@{}}
$\text{num\_tasks} \ge 1$ \\
$\text{num\_tasks} \le \text{num\_cores\_alloc}$
\end{tabular} \\

\bottomrule
\end{tabular}
\end{table}
\begin{table}[t]
\tiny
\centering
\caption{Each dataset is generated from analytical performance models defining the ground-truth \(Y\). Across all datasets, \sysname recommends configurations that only deviates from the ground truth by $<1\%$.
}
\label{tab:syn-mape}
\resizebox{\columnwidth}{!}{
\begin{tabular}{|p{1.5cm}|p{0.5cm}|p{4cm}|}
\hline
\textbf{Dataset} & Error (\%) & \textbf{Description} \\
\hline
{MILC (MPI\_Allreduce) \#Samples: 4749} & \tablecellhl{0.109} & Inputs: number of processes $p$; Output: runtime $t(p)$; defined by $t(p) = 6.3\times10^{-6}\log_2(p)$;  $p \in [64, 131072]$. \\
\hline
{HOMME MPI\_Reduce (small) \#Samples: 4749}  &  \tablecellhl{0.001}& Inputs: $p$; Output: runtime $t(p) = 0.026 + 2.53\times10^{-6}\sqrt{p} + 1.24\times10^{-12}p^3$; $p \in [64, 131072]$. \\
\hline
{HOMME MPI\_Reduce (large) \#Samples: 4749} & \tablecellhl{0.046} & Inputs: $p$; Output: runtime $t(p) = 2.60\times10^{-2}\sqrt{p} + 1.17\times10^{-12}p^3$; $p \in [64, 131072]$. \\
\hline
{HOMME Vlaplace \#Samples: 4749}  & \tablecellhl{0.00} & Inputs: $p$; Output: runtime $t(p) = 0.034 + 1.33\times10^{-10}p^2$; $p \in [64, 131072]$. \\
\hline
{Sweep3D Wavefront \#Samples: 4749} & \tablecellhl{0.014}  & Inputs: $p$; Output: runtime $t(p) = c\sqrt{p}$, where $c$ is a kernel-dependent constant; $p \in [64, 131072]$, $c = 1.0e-6$. \\
\hline
{Hoefler Communication \#Samples: 6174} & \tablecellhl{9.6e-4} & Inputs: grid dimensions $p_x, p_y$ and number of sweeps $n_{\text{sweep}}$; Output: communication time $t_{\text{comm}} = [2(p_x + p_y - 2) + 4(n_{\text{sweep}} - 1)]\,t_{\text{msg}}$, where $t_{\text{msg}}$ is time per message.
\\
\hline
{Roofline Model (Compute--Memory Bound) \#Samples: 10,000} & \tablecellhl{0.00} & Input: operational intensity $I$ (FLOPs per byte); Output: achievable performance $y = \min(P_{\text{peak}}, B I)$, where $P_{\text{peak}}$ is peak FLOP/s and $B$ is memory bandwidth (bytes/s); $I > 0$, $P_{\text{peak}} = 1e12$, $B =  5e10$. \\
\hline
{Amdahl's Law Scaling \#Samples: 7287}  &  \tablecellhl{0.00}  & Input: number of processes $p$; Output: theoretical speedup $y = \frac{1}{f + \frac{1-f}{p}}$, where $f$ is the serial fraction of the workload; $p \ge 1$, $f = 0.1$. \\
\hline
{Roofline-Derived GPU Model 
\#Samples: 10,000} &\tablecellhl{0.783}  & Input: operational intensity $I$; Output: performance $y = \frac{I}{1/B + 1/P}$, where $B$ is memory bandwidth and $P$ is peak compute throughput; $I > 0$, $B = 8e10 $, $P = 2e12 $. \\
\hline
{Basic Linear (15 Features) \#Samples: 6,000} &\tablecellhl{0.076}  & Inputs: $X_1, X_2, ..., X_{15} $; Output: $y = 2X_{14}-1.5X_8+0.5X_6$. \\
\hline
\end{tabular}
}
\end{table}
\textbf{Synthetic Benchmarking.} In addition to the literature-supported validation for the real-world datasets, we also validate the quality of configurations generated by \sysname using synthetic benchmarks with analytically defined performance models, where the ground-truth relationship between inputs $X$ and outcomes $Y$ is known.
For each benchmark, \sysname is given target outcomes $Y$ and tasked with generating input configurations $X$ whose predicted outcomes reproduce the analytically specified $X \rightarrow Y$ mapping.
This controlled setting isolates the generative engine's ability to reconstruct complex functional relationships without confounding effects from noise or data sparsity, following established benchmarking practices~\cite{matsubara2024rethinkingsymbolicregressiondatasets, calotoiu2017automatic}.
Accuracy is measured using per-sample Absolute Percentage Error (APE),
$\text{APE} = \frac{\lvert \text{Actual} - \text{Forecast} \rvert}{\max(\lvert \text{Actual} \rvert, 10^{-8})} \times 100$,
where lower APE indicates closer agreement with both the functional form and scaling magnitude of the analytical model.

\textbf{Datasets.} 
Table~\ref{tab:syn-mape} lists the datasets and the analytical equations used in this step. The validation suite includes \gls{hpc} benchmarks---MILC, HOMME–MPI\_Reduce (small and large), HOMME–Vlaplace, and Sweep3D~\cite{calotoiu2013using}---alongside a communication model based on Hoefler's equation~\cite{calotoiu2017automatic}. These represent a variety of workloads: MILC and HOMME capture communication-bound behaviors, Sweep3D represents memory-bound kernels, and Hoefler's model formalizes collective communication costs. We also include the Roofline model~\cite{Roofline_CACM09,yang2019hierarchical}, Amdahl's Law~\cite{amdahl1967validity}, and a baseline linear dataset. Together, these models cover major HPC scaling patterns—including linear, polynomial, logarithmic, and composite functions (e.g., $\sqrt{p}$, $p^3$)—to verify if \sysname can reconstruct the deterministic relationships between configuration parameters and performance.

\textbf{Observations.} 
Table~\ref{tab:syn-mape} summarizes the results. Across all benchmarks, \sysname achieves minimum APE values below 1\%, confirming that 
\sysname accurately reproduces analytical scaling relationships when expressed through the \reco query. Low APE scores on simple functional forms demonstrate that \sysname learns correct input–output mappings; slight deviations in composite kernels (e.g., HOMME's mixed terms) reflect expected increases in model complexity, which \sysname handles automatically. \sysname identified Random Forest as the surrogate for most datasets, while the MILC MPI\_Allreduce and Roofline-derived GPU models were better fitted by Multi-Layer Perceptrons (MLPs), indicating non-linear relationships, which matches with the known literature~\cite{calotoiu2017automatic}.

\begin{figure}[t]
    \centering
    \includegraphics[width=\columnwidth]{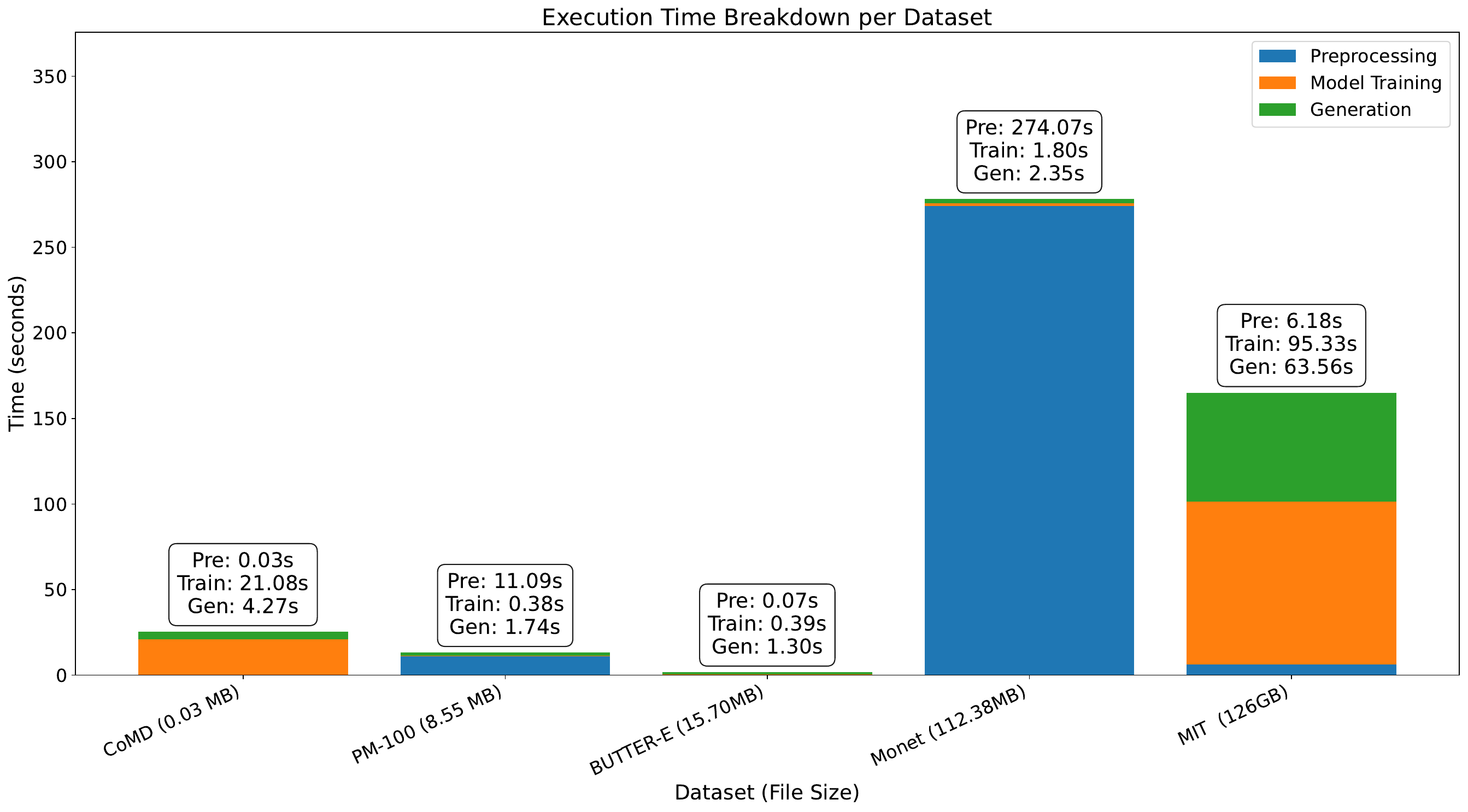}
    \caption{Overhead of \sysname.}
    \label{fig:appendix_timing}
\end{figure}
\begin{table}[t]
\centering
\caption{Comparison of \sysname with different Autotuning libraries, generative and predictive modeling. Predictive modeling has no search criteria, they predict target on given configuration, opposite to our work}
\label{tab:comparison_autotunner}
\resizebox{\columnwidth}{!}{
\begin{tabular}{|p{1.5cm}|p{1cm}|p{1cm}|p{1cm}|p{1cm}|p{1cm}|p{1cm}|p{1cm}|p{1cm}|p{1cm}|p{1cm}|p{1cm}|p{1.5cm}|}
\hline
\textbf{Features}
  & \multicolumn{6}{c|}{\cellcolor{blue!20}\textbf{Autotuner}}
  & \multicolumn{4}{c|}{\cellcolor{green!20}\textbf{Generative}}
  & \multicolumn{1}{c|}{\cellcolor{orange!20}\textbf{Predictive}}
  & \multicolumn{1}{c|}{\cellcolor{violet!20}\textbf{Decision}}
\\ \hline
  & ytopt & Open-Tuner & Harmony & HYPERF & GPTune & BoFire & GReaT & CTGAN & GAN & Diffusion & ModelX & \tablecellbl{\textbf{\sysname}} \\
\hline
Search Criteria
  & \cell{Min, Max} & \cell{Min, Max} & \cell{Min, Max} & \cell{Min, Max} & \cell{Min, Max} & \cell{Min, Max}
  & \cell{Min, Max, Range} & \cell{Min, Max} & \cell{Cat., Cont.} & \cell{Cat., Cont.}
  &  \ding{56}
  & \tablecellbl{\cell{Cat, Min, Max, Range}} \\
\hline
\cell{\reco}
  & \ding{52} & \ding{52} & \ding{52} & \ding{52} & \ding{52} & \ding{52}
  & \ding{52} & \ding{52} & \ding{52} & \ding{52}
  & \ding{56}
  & \tablecellbl{\ding{52}} \\
\hline
\cell{\modify}
  & \ding{56} & \ding{56} & \ding{56} & \ding{56} & \ding{56} & \ding{56}
  & \ding{52} & \ding{52} & \ding{56} & \ding{56}
  & \ding{56}
  & \tablecellbl{\ding{52}} \\
\hline
\cell{\whatif}
  & \ding{56} & \ding{56} & \ding{56} & \ding{56} & \ding{56} & \ding{56}
  & \ding{52} & \ding{52} & \ding{56} & \ding{56}
  & \ding{56}
  & \tablecellbl{\ding{52}} \\
\hline
Query-Driven Interface
  & \ding{56} & \ding{56} & \ding{56} & \ding{56} & \ding{56} & \ding{56}
  & \ding{52} & \ding{56} & \ding{56} & \ding{56}
  & \ding{56}
  & \tablecellbl{\ding{52}} \\
\hline
Reliability Assessment
  & \ding{56} & \ding{56} & \ding{56} & \ding{56} & \ding{56} & \ding{56}
  & \ding{52} & \ding{52} & \ding{52} & \ding{52}
  & \ding{52}
  & \tablecellbl{\ding{52}} \\
\hline
Interpretable Explanation
  & \ding{56} & \ding{56} & \ding{56} & \ding{56} & \ding{56} & \ding{56}
  & \ding{52} & \ding{52} & \ding{52} & \ding{52}
  & \ding{52}
  & \tablecellbl{\ding{52}} \\
\hline
Satisficing Solutions
  & \ding{56} & \ding{56} & \ding{56} & \ding{56} & \ding{56} & \ding{56}
  & \ding{52} & \ding{56} & \ding{56} & \ding{56}
  & \ding{56}
  & \tablecellbl{\ding{52}} \\
\hline
\end{tabular}
}
\end{table}

\end{document}